\documentclass{article}

\PassOptionsToPackage{numbers,compress}{natbib}

\usepackage[position,preprint]{neurips_2026}

\usepackage{amsmath,amssymb,amsfonts}
\usepackage{graphicx}
\usepackage{textcomp}
\usepackage{xcolor}
\def\BibTeX{{\rm B\kern-.05em{\sc i\kern-.025em b}\kern-.08em
    T\kern-.1667em\lower.7ex\hbox{E}\kern-.125emX}}

\usepackage[all]{nowidow}
\usepackage{listings}
\usepackage{url}
\usepackage{makecell}
\usepackage{hyperref}
\usepackage{adjustbox}
\usepackage{xspace}
\usepackage{multirow}
\usepackage{tikz}
\usepackage{ctable}
\usepackage{color}

\usepackage{algorithm}

\usepackage{algpseudocode}

\usepackage{cleveref}
\usepackage{acronym}
\usepackage{fancybox}
\usepackage{verbatim}
\usepackage[normalem]{ulem}
\usepackage{float}
\usepackage{newfloat}
\usepackage{mdframed}

\usepackage{enumitem}
\usepackage{booktabs}
\usepackage{pifont}
\usepackage[flushleft]{threeparttable}
\usepackage{tabularray}

\usepackage[most]{tcolorbox}

\usepackage{rotating}

\usepackage{colortbl}

\usepackage{pdfrender}

\newcommand{\TODO}[1]{}

\newcommand{\npapers}{59\xspace}
\newcommand{\nsystems}{21\xspace}
\newcommand{\nplugins}{26\xspace}

\newcommand{\ahi}{AHI\xspace}
\newcommand{\ahisec}{AHI security\xspace}
\newcommand{\datayear}{April 2026\xspace}

\newcommand{\claudecode}{Claude Code\xspace}
\newcommand{\codex}{Codex\xspace}
\newcommand{\cursorsys}{Cursor\xspace}
\newcommand{\aider}{Aider\xspace}
\newcommand{\devin}{Devin\xspace}
\newcommand{\manus}{Manus\xspace}

\newcommand{\ticon}[1]{\raisebox{-.2em}{\includegraphics[height=0.9em]{figures/icons/#1}}}

\newcommand{\iconDetEngine}{\ticon{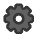}}    
\newcommand{\iconLLMChecker}{\ticon{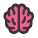}}         
\newcommand{\iconLLMSelf}{\ticon{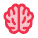}}       
\newcommand{\iconFwHook}{\ticon{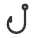}}             
\newcommand{\iconSysEnf}{\ticon{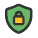}}         
\newcommand{\iconLLMFw}{\ticon{LLM-as-checker.pdf}\kern1pt\ticon{framework-hook.pdf}} 

\newcommand{\iconStatic}{\ticon{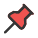}}       
\newcommand{\iconDynamic}{\ticon{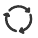}}     
\newcommand{\iconSession}{\ticon{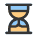}}     

\newcommand{\iconCoarse}{$\bullet$}            
\newcommand{\iconMedium}{$\bullet\bullet$}     
\newcommand{\iconFine}{$\bullet\bullet\bullet$} 
\newcommand{\iconDynGran}{\ticon{dynamic.pdf}}     

\newcommand{\iconLattice}{$\Diamond$}          
\newcommand{\iconCapBased}{\ticon{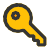}} 
\newcommand{\iconBinaryTrust}{\ticon{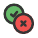}} 
\newcommand{\iconMultiLevel}{\ticon{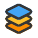}} 

\newcommand{\iconDetTaint}{\ticon{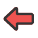}} 
\newcommand{\iconLLMReason}{\ticon{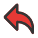}}     
\newcommand{\iconManual}{\ticon{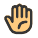}}                
\newcommand{\iconSelective}{\ticon{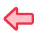}}          

\newcommand{\iconSecExpert}{\ticon{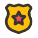}}  
\newcommand{\iconEndUser}{\ticon{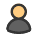}}           
\newcommand{\iconDeveloper}{\ticon{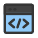}}        

\newcommand{\iconIFC}{\ticon{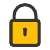}}              
\newcommand{\iconCapCheck}{\ticon{capability-check.pdf}}   
\newcommand{\iconIsolated}{\ticon{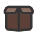}} 
\newcommand{\iconFilter}{\ticon{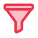}}     
\newcommand{\iconInputFilter}{\ticon{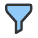}} 
\newcommand{\iconFormalVerif}{\ticon{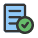}} 

\newcommand{\iconBinary}{\ticon{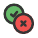}}               
\newcommand{\iconBinaryMem}{\ticon{binary.pdf}\kern1pt\ticon{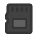}}     
\newcommand{\iconBinaryFb}{\ticon{binary.pdf}\kern1pt\ticon{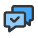}}    
\newcommand{\iconThreeWay}{\ticon{binary.pdf}\kern1pt\ticon{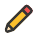}}        
\newcommand{\iconAdvisory}{\ticon{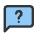}}           

\newcommand{\iconFramework}{\ticon{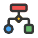}}         
\newcommand{\iconOS}{\ticon{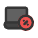}}                       
\newcommand{\iconVM}{\ticon{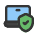}}                       
\newcommand{\iconCloud}{\ticon{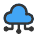}}                 
\newcommand{\iconApp}{\ticon{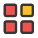}}             
\newcommand{\iconNone}{\ticon{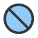}}                   

\newcommand{\iconOpen}{\tikz[baseline=-.3em]{\fill[green!70!black] (0,0) circle (0.35em);}}    
\newcommand{\iconClosed}{\tikz[baseline=-.3em]{\fill[red!80!black] (0,0) circle (0.35em);}}    
\newcommand{\iconPartial}{\tikz[baseline=-.3em]{\fill[yellow!70!black] (0,0) circle (0.35em);}} 
\newcommand{\iconTiered}{\textcolor{orange}{$\blacklozenge$}}    
\newcommand{\iconConfigurable}{\ticon{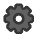}}             

\newcommand{\iconImmutable}{\ticon{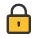}}              
\newcommand{\iconExpandOnly}{\ticon{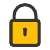}}        
\newcommand{\iconFlexible}{$\circlearrowleft$}         

\newcommand{\iconSinglePrompt}{\ticon{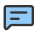}}  
\newcommand{\iconDynRules}{\ticon{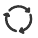}}            
\newcommand{\iconBehavHist}{\ticon{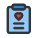}}        

\newcommand{\iconSemAlign}{\ticon{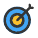}}       
\newcommand{\iconRuleMatch}{\ticon{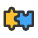}}              
\newcommand{\iconReExec}{\ticon{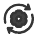}}       
\newcommand{\iconBehavAnom}{\ticon{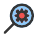}}      

\newcommand{\iconPerStep}{\ticon{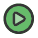}}            
\newcommand{\iconPerStepCont}{\ticon{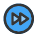}}   
\newcommand{\iconBatch}{\ticon{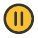}}                 
\newcommand{\iconContinuous}{\ticon{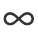}}       

\newcommand{\iconInformed}{$\odot$}            
\newcommand{\iconNoAware}{$\emptyset$}         
\newcommand{\iconIndirect}{$\oslash$}          

\newcommand{\iconCmd}{\ticon{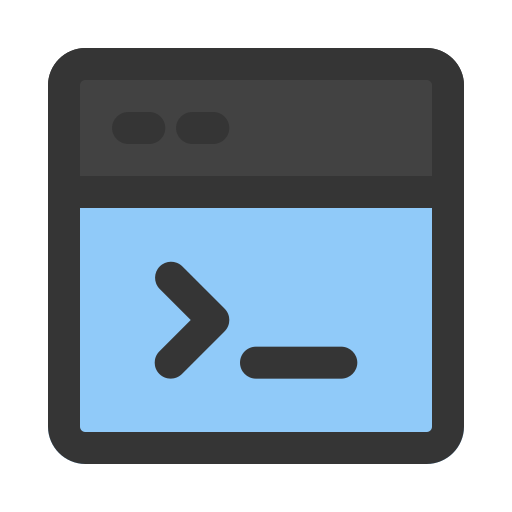}}         
\newcommand{\iconMCP}{\ticon{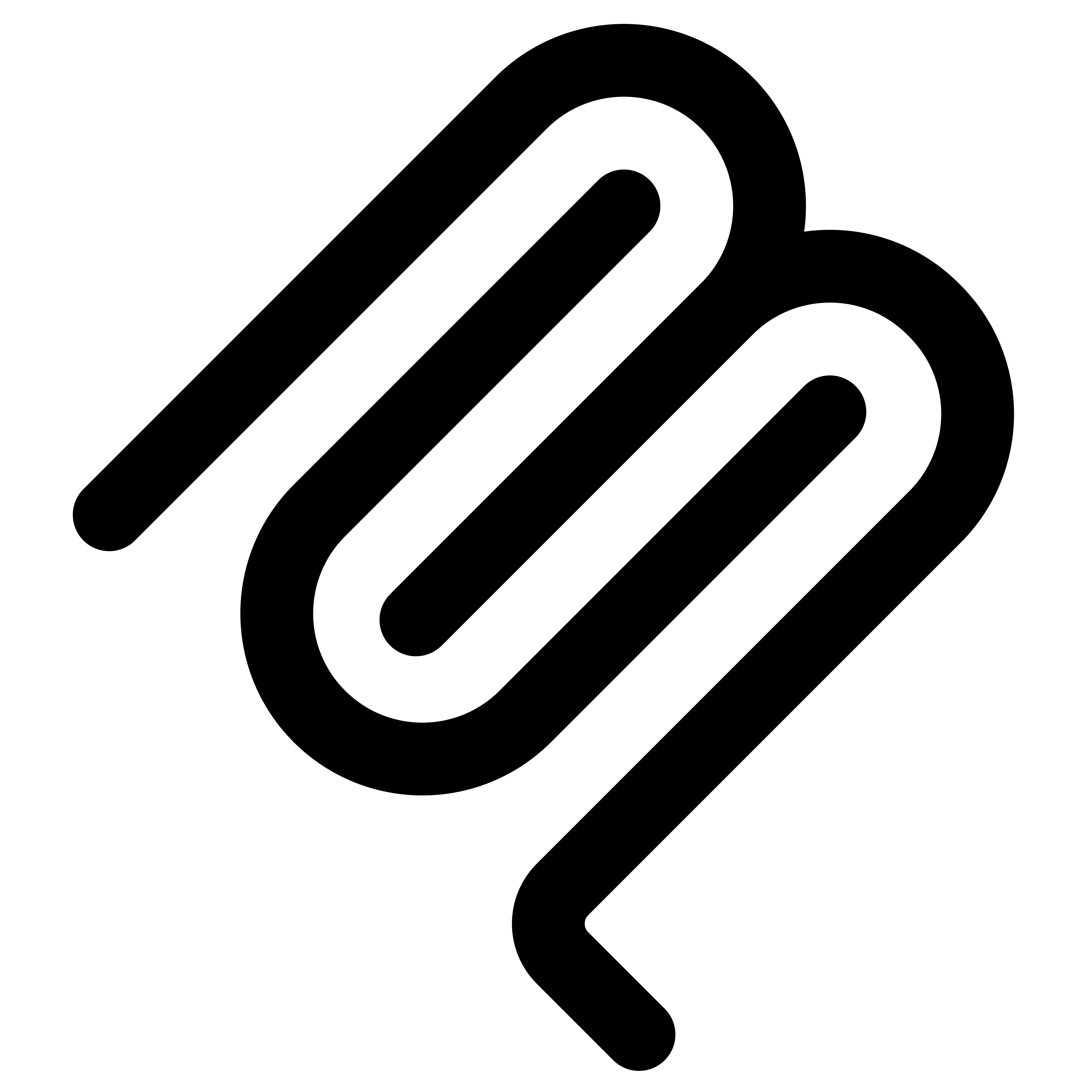}}         
\newcommand{\iconAPI}{\ticon{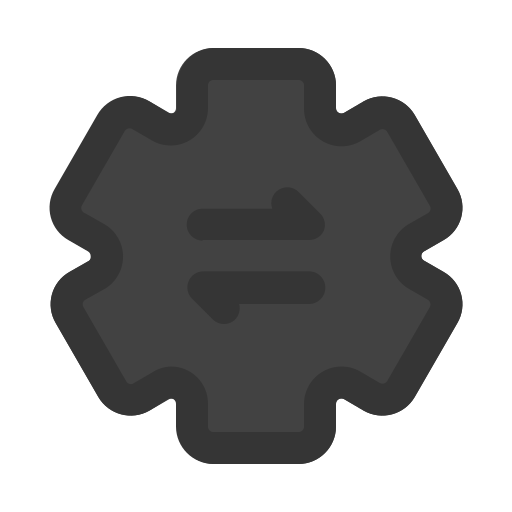}}         
\newcommand{\iconDir}{\ticon{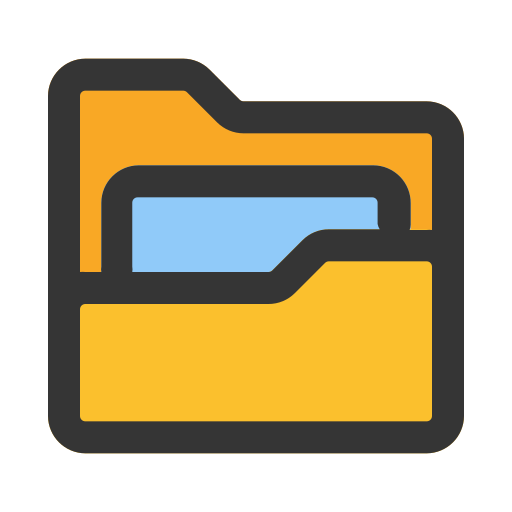}}         
\newcommand{\iconFile}{\ticon{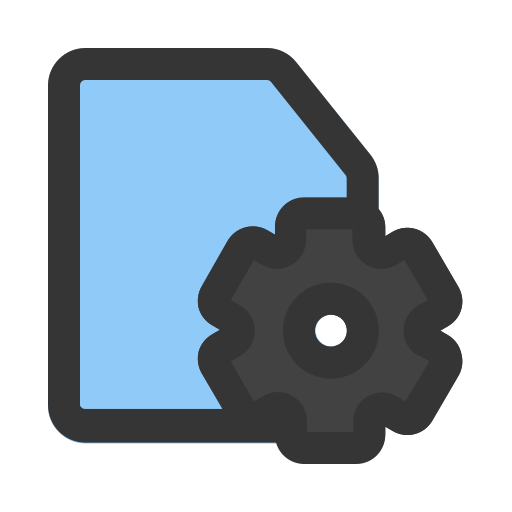}}       
\newcommand{\iconNet}{\ticon{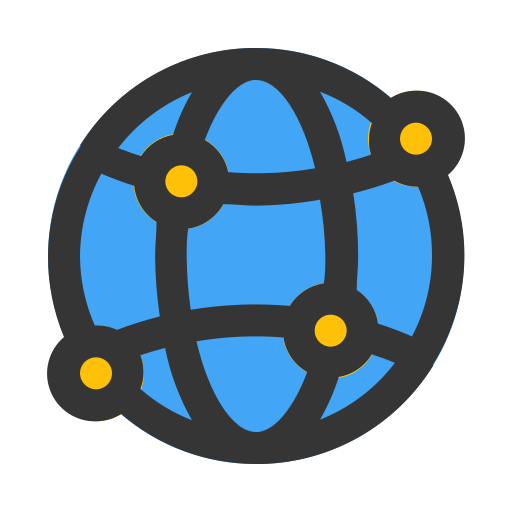}}         
\newcommand{\iconTool}{\ticon{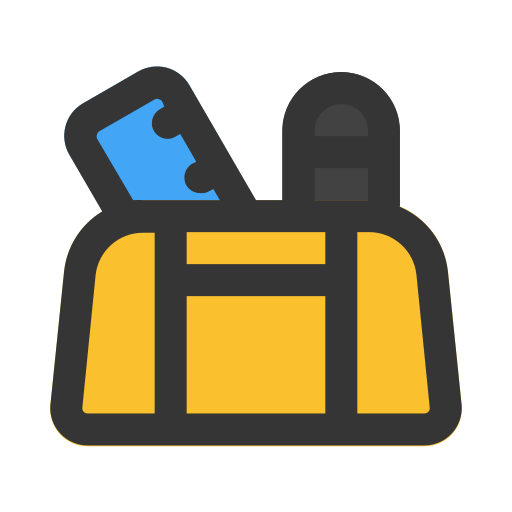}}       
\newcommand{\iconEnv}{\ticon{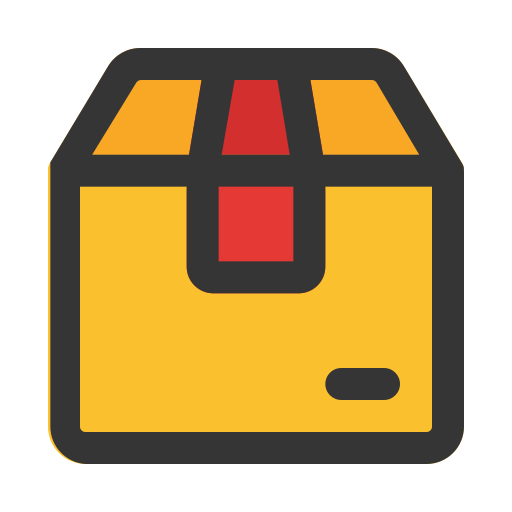}}         
\newcommand{\iconRepo}{\ticon{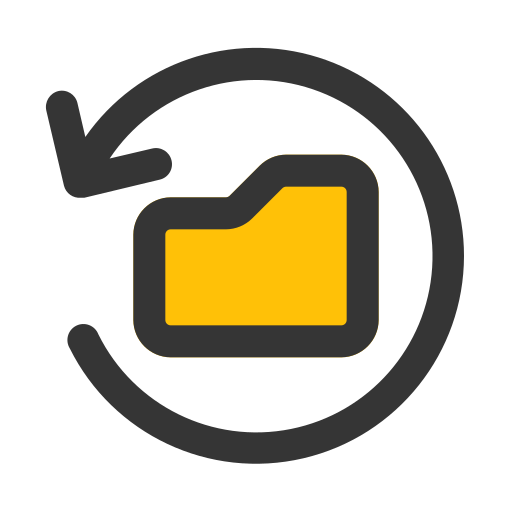}}       
\newcommand{\iconCode}{\ticon{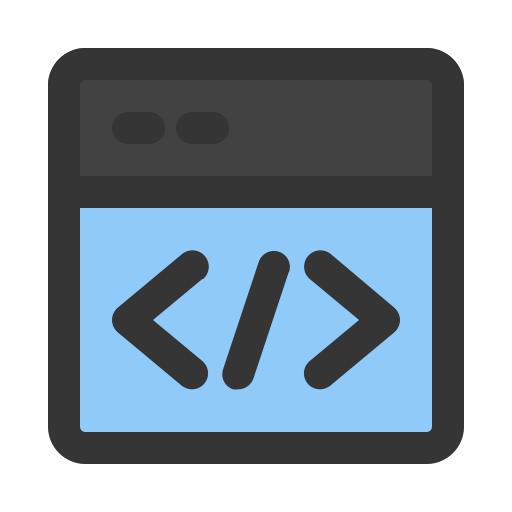}}       
\newcommand{\iconConfig}{\ticon{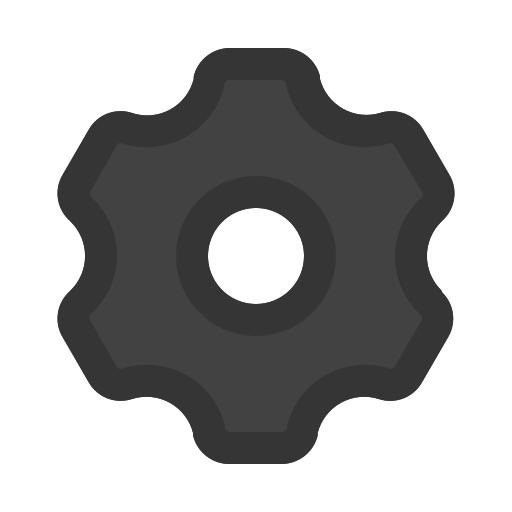}}   
\newcommand{\iconPlan}{\ticon{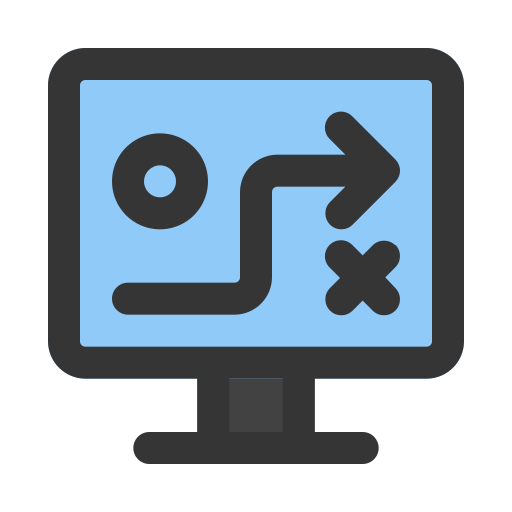}}       
\newcommand{\iconPR}{\ticon{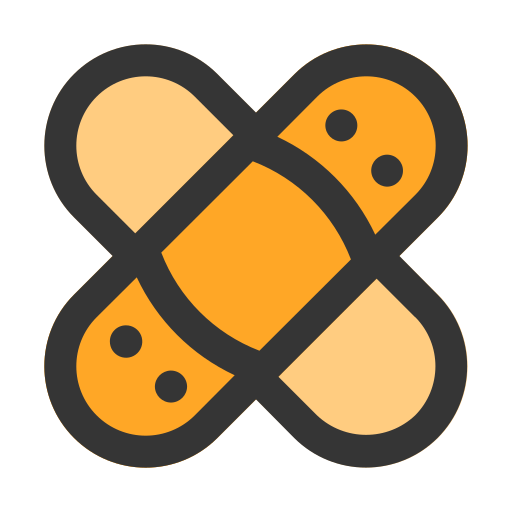}}           
\newcommand{\iconData}{\ticon{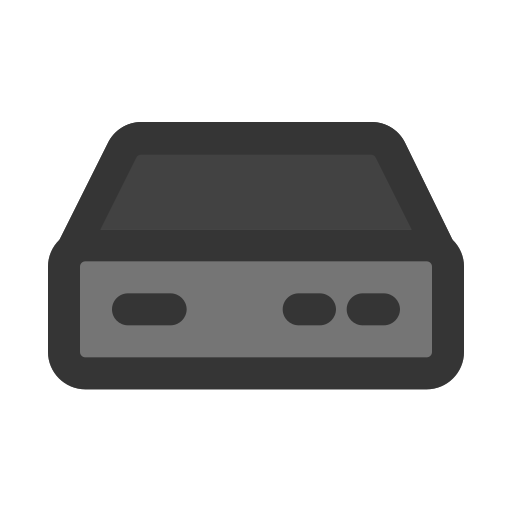}}       
\newcommand{\iconAgent}{\ticon{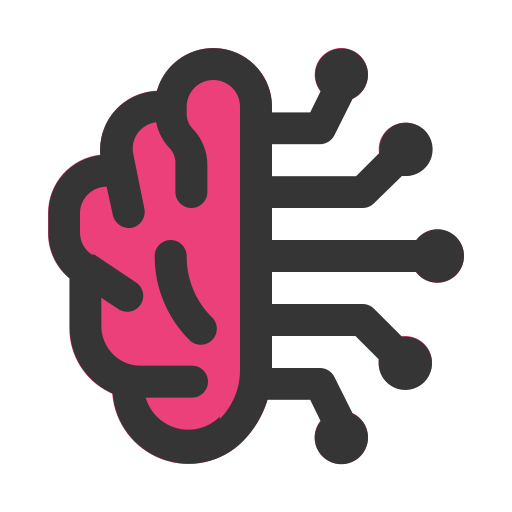}}     
\newcommand{\iconAdmin}{\ticon{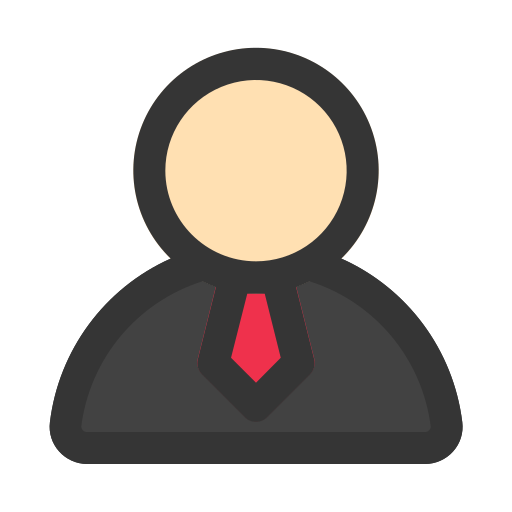}}     
\newcommand{\iconLLM}{\ticon{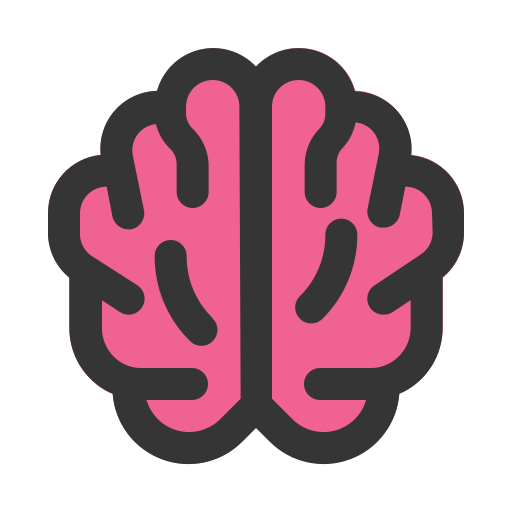}}         
\newcommand{\iconAuth}{\ticon{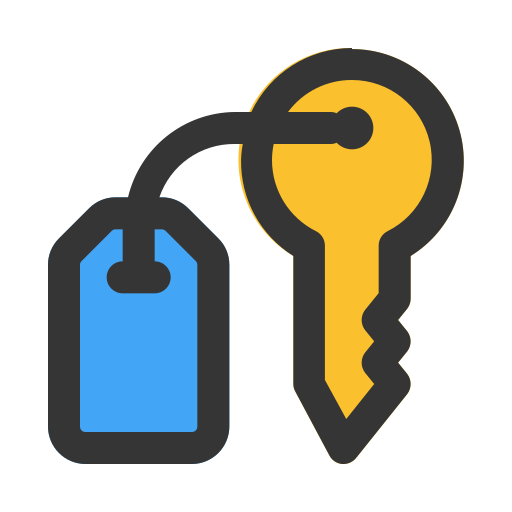}}       
\newcommand{\iconReview}{\ticon{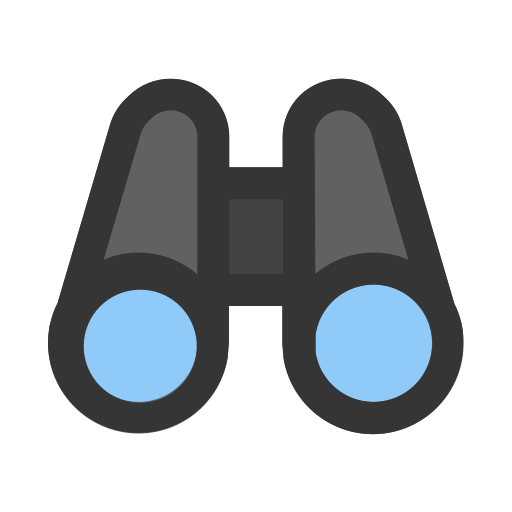}}   
\newcommand{\iconDiff}{\ticon{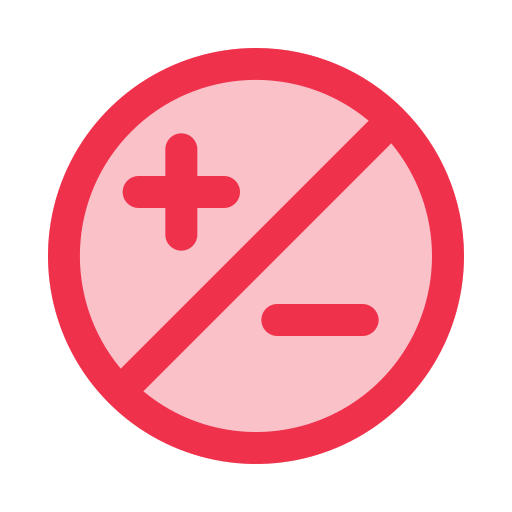}}       
\newcommand{\iconGUI}{\ticon{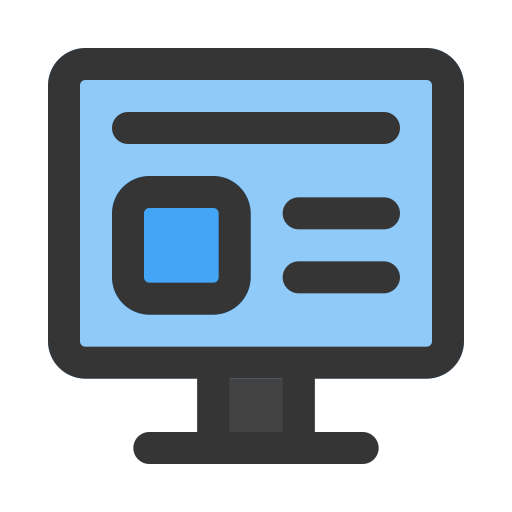}}         
\newcommand{\iconProxy}{\ticon{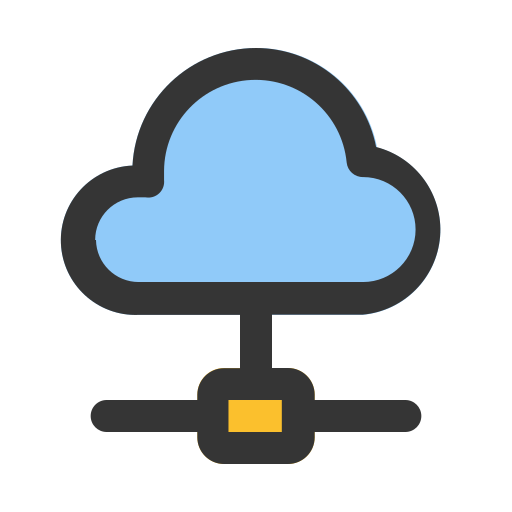}}     

\newcolumntype{M}[1]{>{\centering\arraybackslash}p{#1}}
\newcommand{\sssec}[1]{\vspace*{0.05in}\noindent\textbf{#1}}

\usepackage{contour}
\contourlength{0.4pt} 
\contournumber{20}

\definecolor{mygray}{rgb}{0.95,0.95,0.95}

\definecolor{rowPaper}{RGB}{225,237,250}   
\definecolor{rowSystem}{RGB}{225,245,225}  
\definecolor{rowPlugin}{RGB}{255,232,210}  

\newcommand{\ratebar}[1]{%
  \begin{tikzpicture}[baseline=-0.3ex]
    \draw[gray!55, line width=0.25pt] (0,0) rectangle (1.0,0.20);
    \fill[green!50!black] (0,0) rectangle ({#1/100*1.0},0.20);
  \end{tikzpicture}%
}

\newcommand{\iconScopeCfg}{\raisebox{-0.3ex}{\includegraphics[height=1.4em]{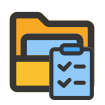}}}
\newcommand{\iconRuntimeApproval}{\raisebox{-0.3ex}{\includegraphics[height=1.4em]{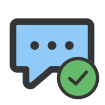}}}
\newcommand{\iconPolicySpec}{\raisebox{-0.3ex}{\includegraphics[height=1.4em]{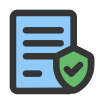}}}
\newcommand{\iconTrustLabel}{\raisebox{-0.3ex}{\includegraphics[height=1.4em]{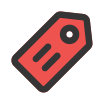}}}
\newcommand{\iconIntentAnchor}{\raisebox{-0.3ex}{\includegraphics[height=1.4em]{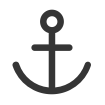}}}

\newtcolorbox[auto counter, number format=\Alph]{study}[2][]{
    detach title,
    before upper={\tcbtitle\quad},
    colback=mygray,
    enhanced,
    fonttitle=\bfseries\itshape,
    breakable,
    colframe=white,
    left=0pt,right=0pt,top=0pt,bottom=0pt,
    title={Case study~\thetcbcounter. #1.},
    sharp corners=northwest, 
    sharp corners=southwest, 
    coltitle=black,
    colbacktitle=mygray,
    boxrule=0pt,
    frame hidden,
    leftrule=1pt, toprule=0pt, rightrule=0pt, bottomrule=0pt,
    borderline west={1pt}{0pt}{black},
    #2,
}

\newcolumntype{L}[1]{>{\raggedright\arraybackslash}p{#1}}

\newcounter{takeaway}
\newcommand{\takeaways}[1]{
\vspace{1pt}
\noindent
\begin{tcolorbox}[ enhanced,
    breakable,
    boxrule=1pt, 
    arc=4pt,
    left=2pt,
    right=2pt,
    bottom=2pt,
    top=2pt,
    colback=gray!4,        
    colframe=gray!1!black,
    drop shadow=black!50!white, 
    rounded corners]
\noindent
\refstepcounter{takeaway}
\textbf{Takeaway \Roman{takeaway}.} \small
{#1}
\end{tcolorbox}
}

\newcounter{openproblem}
\newcommand{\openproblem}[1]{
\vspace{1pt}
\noindent
\begin{tcolorbox}[ enhanced,
    breakable,
    boxrule=0.8pt,
    arc=4pt,
    left=2pt,
    right=2pt,
    bottom=2pt,
    top=2pt,
    colback=orange!5,
    colframe=orange!60!black,
    borderline={0.8pt}{0pt}{orange!60!black, dashed},
    rounded corners]
\noindent
\refstepcounter{openproblem}
\textbf{Open Problem \Roman{openproblem}.} \small
{#1}
\end{tcolorbox}
}

\title{Reframing LLM Agent Security as an Agent--Human Interaction Problem}

\author{%
  Peiran Wang \\
  UCLA \\
  \texttt{peiranwang@ucla.edu} \\
  \And
  Ying Li \\
  UCLA \\
  \texttt{ying.li@ucla.edu} \\
  \And
  Yuan Tian \\
  UCLA \\
  \texttt{yuant@ucla.edu} \\
}

\begin{document}

\maketitle

\begin{abstract}
We argue that LLM agent security is fundamentally an agent-human interaction (AHI) problem, not a purely algorithmic one.
To substantiate this position, we conduct a systematic analysis of \npapers academic papers, \nsystems production agent systems, and \nplugins security plugins as of \datayear.
Our analysis reveals a striking pattern: the three widely deployed human-centric security mechanisms (policy specification, runtime approval, and scope configuration) dominate industry practice, each adopted by at least 14 of \nsystems systems (14, 15, and 16, respectively), while the categories most heavily studied in academia (intent anchoring and trust labeling) see zero production deployment.
Yet current human participation mechanisms are far from satisfactory: they suffer from a fundamental trade-off between cognitive burden and security guarantees, leaving users caught between approval fatigue and uncontrolled agent autonomy.
We make three contributions.
First, through a systematic comparison of LLM-based and human-based intent alignment, we argue that human participation in agent security decisions is indispensable given current capabilities.
Second, we quantify a pronounced industry--academia mismatch: the security mechanisms that practitioners actually deploy receive scant research attention, while the approaches that researchers favor remain undeployed.
Third, we propose a three-direction research agenda and call for \ahisec to be recognized as a first-class research citizen, one that demands its own design principles, evaluation methods, and theoretical foundations.
\end{abstract}

\section{Introduction}
\label{sec:intro}

LLM-based agents are rapidly acquiring capabilities that carry real-world consequences: executing arbitrary shell commands, modifying codebases, orchestrating multi-step workflows through external APIs, and even browsing the web on a user's behalf~\cite{anthropic2025claudecode, christodorescu2025systems}.
Systems such as \claudecode, \codex, \cursorsys, \devin, and \manus now routinely perform actions that, if misdirected by prompt injection, hallucination, or misconfiguration, can compromise confidentiality, integrity, and availability of user data and infrastructure.
The security stakes of these agentic systems are no longer hypothetical; they are operational.
\textbf{We argue that LLM agent security is fundamentally an agent-human interaction (\ahi) problem, not a purely algorithmic one.}
This position is grounded in an empirical observation: every one of the \nsystems production agent systems we analyzed keeps humans in the security loop, relying on some form of user approval, scope configuration, or policy specification to bound agent behavior.
No deployed system trusts an LLM alone to determine whether an action is safe.

Despite this universal industry practice, the academic security community has largely pursued a different trajectory.
The dominant research agenda targets fully automated defenses (prompt injection detection~\cite{wang2026landscape}, formal verification of agent plans, and LLM-as-judge intent validation~\cite{zheng2023judging}) with the implicit assumption that sufficiently capable models will eventually eliminate the need for human oversight.
Our data tells a different story.
Among the five security mechanism categories we identify, intent anchoring, the category that most closely embodies this automated ideal, has a production adoption rate of exactly zero out of \nsystems systems.
Meanwhile, the human-centric categories policy specification, runtime approval, and scope configuration are each deployed by at least 14 of \nsystems systems (14, 15, and 16, respectively), representing the overwhelming majority of real-world security infrastructure.
Yet the current state of human participation is deeply unsatisfying.
Users suffer from approval fatigue when confronted with repeated confirmation dialogs~\cite{feng2025regulatory}; scope boundaries configured once at session start fail to adapt to evolving tasks; and policy languages remain inaccessible to non-expert users.
The problem is not a lack of human involvement; rather, it is that current involvement is poorly designed, under-studied, and lacks theoretical grounding.

This paper makes three contributions in support of our position.
\textbf{First}, we present a systematic comparison of two paths to intent alignment, namely LLM-based automated verification versus human-in-the-loop confirmation, and argue that current LLM capabilities are insufficient to close the intent alignment gap without human participation.
Automated approaches face fundamental limitations: they cannot reliably distinguish adversarial instructions from legitimate ones in open-ended tool-use settings, and they lack access to the user's evolving intent, contextual preferences, and risk tolerance.
Humans, despite their cognitive limitations, remain the only available ground-truth oracle for whether an agent action aligns with the user's actual goals.
\textbf{Second}, drawing on a systematic analysis of \npapers papers, \nsystems production systems, and \nplugins security plugins (as of \datayear), we quantify a pronounced mismatch between industry deployment and academic research focus.
Human-centric mechanisms that dominate production receive disproportionately little research attention, while automated approaches that dominate publication venues see no deployment.
This gap represents both a missed opportunity for researchers and a risk for practitioners who lack evidence-based design guidance.
\textbf{Third}, we propose a three-direction research agenda, encompassing \emph{cognitive modeling} of user security decisions, \emph{adaptive interaction design} that calibrates human involvement to risk, and \emph{hybrid architectures} that combine automated pre-filtering with targeted human judgment, and call for \ahisec to be elevated to a first-class research citizen within the security and AI communities.

The remainder of this paper is organized as follows.
\Cref{sec:sok} introduces our \ahi{} framework, develops the two-path comparison, and quantifies the industry--academia mismatch.
\Cref{sec:position} presents our core position through a thesis--antithesis--synthesis structure.
\Cref{sec:alt-views} addresses alternative views.
\Cref{sec:agenda} details the proposed research agenda, and \Cref{sec:conclusion} concludes.


\section{Background: Intent Alignment and Two Paths}
\label{sec:sok}

\subsection{The Core Challenge: Intent Alignment}
\label{sec:bg-intent}

LLM agents operate in open-ended, dynamic environments: they browse the web, invoke APIs, modify files, and interact with third-party services.
Because the space of possible actions is vast and context-dependent, a user cannot fully pre-specify every permission an agent may need before a task begins.
New sub-goals emerge during execution, external state changes unpredictably, and the agent must replan on the fly~\citep{yao2022react, huang2024understanding}.
This dynamism rules out any static, upfront authorization scheme as a complete solution.

Worse still, many security-relevant decisions are \emph{semantic} rather than syntactic.
Deciding whether ``delete all \texttt{.tmp} files in the project directory'' is safe requires understanding the user's project structure and workflow, not merely pattern-matching against a list of forbidden syscalls.
Pure programmatic rules lack the expressiveness to capture such judgments~\citep{tsai2025contextual}.

These two observations, the impossibility of full pre-specification and the inadequacy of rule-based checking, converge on a single core problem: \textbf{how can we judge whether an agent's action aligns with the user's intent?}
We call this the \emph{intent alignment} challenge.
Our SoK data confirms that intent alignment is not one defense category among many; it is the unifying thread.
Task-alignment mechanisms check whether individual actions serve the user's stated goal; policy controls let users encode behavioral constraints; prompt-isolation techniques ensure that the agent's plan derives from trusted instructions rather than injected directives~\citep{wang2026landscape}.
Every defense concept we catalogued ultimately serves intent alignment.

\subsection{Two Complementary Paths}
\label{sec:bg-paths}

Given that intent alignment is the central challenge, existing defenses pursue it through two fundamentally different strategies.

\begin{figure}[t]
  \centering
  \includegraphics[width=\linewidth]{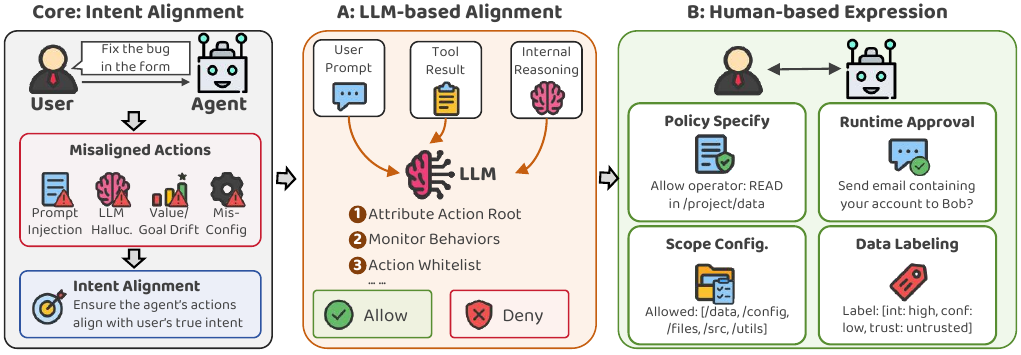}
  \caption{Two complementary paths for intent alignment in LLM agent security. Path~A uses LLMs to infer user intent automatically; Path~B lets humans express intent directly through policies, approvals, and scope configurations.}
  \label{fig:two-paths}
\end{figure}

\sssec{Path~A: LLM-based intent alignment (academia-driven).}
The first strategy delegates intent judgment to another LLM.
A secondary model (or a specialized head of the same model) observes the agent's proposed action together with the original user prompt and decides whether the action is justified.
This family includes \emph{task-alignment} classifiers that score each tool call against the user task~\citep{jia2025task, zhu2025melon}, \emph{information-flow-control} frameworks that use an LLM as judge to detect unauthorized data exfiltration~\citep{debenedetti2025defeating, costa2025securing}, and \emph{intent-anchoring} techniques that treat the user prompt as a semantic anchor against which all subsequent actions are validated~\citep{li2025drift, luo2025agrail, xiang2026architecting}.

Path~A is attractive because it promises full automation: the user issues a task and the system handles everything, including security.
In our SoK corpus, 15 of 41 defense works rely on an LLM for intent alignment~\citep{wang2026landscape}.
Path~A maps primarily to our intent anchoring category, which contains 5 papers but, crucially, \emph{zero} production deployments out of \nsystems systems surveyed.

\sssec{Path~B: Human-based intent expression (industry-driven).}
The second strategy keeps the human in the loop.
Rather than inferring intent, it provides structured channels through which users \emph{directly express, confirm, or constrain} their intent.
These channels correspond to four of our five \ahi{} categories:
\begin{itemize}[noitemsep, topsep=1pt, leftmargin=*]
  \item \iconPolicySpec~\textbf{Policy specification:} Users write rules, ranging from natural-language instructions to formal policies, that prescribe or forbid specific agent behaviors~\citep{shi2025progent, wang2026agentspec}.
  \item \iconRuntimeApproval~\textbf{Runtime approval:} Users approve or reject individual operations at runtime, retaining a veto over high-stakes actions~\citep{wu2024isolategpt, zhong2025rtbas}.
  \item \iconScopeCfg~\textbf{Scope and boundary configuration:} Users configure boundaries (permitted directories, accessible APIs, network ranges) before or during execution~\citep{zhu2025miniscope, yan2025fault}.
  \item \iconTrustLabel~\textbf{Trust and data labeling:} Users annotate data sources or tool outputs with trust labels that inform downstream access control~\citep{debenedetti2025defeating, siddiqui2024permissive}.
\end{itemize}
Path~B is the dominant strategy in production: 9--10 of 41 defense works~\citep{wang2026landscape} and the vast majority of deployed systems rely on explicit human participation.
Among Path~B's four categories, three (policy specification, runtime approval, and scope and boundary configuration) are widely deployed (14--16 of \nsystems systems each), while trust and data labeling, despite being human-centric in design, remains academic-only with zero production adoption: a gap we examine in \S\ref{sec:mismatch}.

\sssec{Relationship between the two paths.}
Paths~A and~B are complementary.
LLM-based judgment can reduce user burden by filtering out clearly safe or clearly dangerous actions, while human-based expression provides a ground-truth anchor that no LLM judge can fully replace.
However, the relationship is \emph{asymmetric}: Path~B is more fundamental, because human intent can only originate from the human.
We develop this argument formally in \Cref{sec:position}.

\subsection{Industry--Academia Mismatch}
\label{sec:mismatch}

The complementarity described above would be unremarkable if both paths received proportionate research attention.
They do not.
Our analysis of \nsystems production systems reveals a stark asymmetry between what industry deploys and what academia studies.

\begin{figure}[t]
\centering
\begin{minipage}[b]{0.6\linewidth}
  \makeatletter\renewcommand{\@captype}{table}\makeatother
  \centering
  \small
  \setlength{\tabcolsep}{3pt}
  \renewcommand{\arraystretch}{1.25}
  \begin{tabular}{@{}clccl@{}}
    \toprule
    & \textbf{Category} & \textbf{Papers} & \textbf{Systems} & \textbf{Rate} \\
    \midrule
    \rowcolor{rowSystem} \iconScopeCfg         & Scope \& Boundary Config. & 4  & 16 & \ratebar{76}~76\% \\
    \rowcolor{rowSystem} \iconRuntimeApproval  & Runtime Approval          & 5  & 15 & \ratebar{71}~71\% \\
    \rowcolor{rowSystem} \iconPolicySpec       & Policy Specification      & 13 & 14 & \ratebar{67}~67\% \\
    \rowcolor{rowPaper}  \iconTrustLabel       & Trust \& Data Labeling    & 8  &  0 & \ratebar{0}~0\%   \\
    \rowcolor{rowPaper}  \iconIntentAnchor     & Intent Anchoring          & 5  &  0 & \ratebar{0}~0\%   \\
    \bottomrule
  \end{tabular}
  \caption{Adoption of \ahi{} defense categories, sorted by industry adoption rate. The \textbf{Rate} column reports the percentage of the \nsystems production systems that ship at least one mechanism in that category. Icons in the leftmost column also serve as the legend for \Cref{fig:burden-guarantee}.}
  \label{tab:adoption}
\end{minipage}\hfill
\begin{minipage}[b]{0.35\linewidth}
  \centering
  \includegraphics[width=\linewidth]{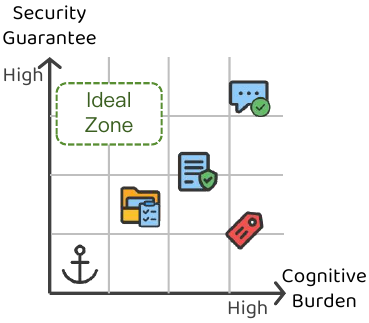}
  \vspace{-2em}
  \caption{Qualitative positioning of the 5 \ahi{} categories on the cognitive burden vs.\ security guarantee plane. 
  }
  \label{fig:burden-guarantee}
\end{minipage}
\end{figure}

\sssec{Industry overwhelmingly chooses human-centric mechanisms.}
\Cref{tab:adoption} summarizes adoption across the five \ahi{} categories.
Runtime approval appears in 15 of \nsystems systems; scope configuration in 16; policy specification in 14.
These are not niche features; they are the \emph{default} security architecture of systems such as \claudecode, \codex, \cursorsys, \devin, and \manus.
By contrast, intent anchoring has \emph{zero} production deployments.
Trust labeling, despite 8~academic papers, likewise has zero adoption in production.

\sssec{Academia still pursues full automation.}
Academic research gravitates toward Path~A.
Five recent papers target intent anchoring, seeking LLM-based mechanisms that can replace human judgment entirely---yet none has been deployed in production.
Policy research, meanwhile, pursues increasingly formal domain-specific languages (DSLs)~\citep{shi2025progent, wang2026agentspec} that achieve strong theoretical guarantees but assume a level of specification skill that ordinary developers, let alone end users, do not possess.
Across our corpus, remarkably little attention is paid to the cognitive and practical constraints of the humans who must actually configure, approve, and oversee agent behavior.

\sssec{Key observation.}
This mismatch is \emph{not} a symptom of industry laziness or technical conservatism.
Production teams at leading AI companies have independently converged on human-centric mechanisms because, through extensive deployment experience, they have discovered that human participation provides a more reliable foundation for intent alignment than any purely automated alternative available today.
When 15 of \nsystems systems implement runtime approval and zero implement intent anchoring, the signal is unambiguous.
Academia must take this signal seriously: the path forward is not to engineer the human out of the loop, but to make the human's role in the loop as effective as possible.
We formalize this argument in the next section.

\section{The Core Position: Security as an Agent--Human Interaction Problem}
\label{sec:position}

We now present our central argument in three stages: we first argue that human
participation is more fundamental than LLM-based automation for intent alignment
(\S\ref{sec:pos-thesis}), then show that current human participation mechanisms
are deeply flawed (\S\ref{sec:pos-antithesis}), and finally synthesize these
observations into our core position (\S\ref{sec:pos-synthesis}).

\subsection{Why Human Participation Is More Fundamental}
\label{sec:pos-thesis}

We advance five reasons, each building on the previous, for why human
participation cannot be engineered away.

\noindent\textbf{Reason 1: LLM alignment is built on user prompts, but user prompts are imprecise.}
LLM-based intent alignment treats the user's prompt as ground truth against
which agent actions are validated. But user prompts are rarely precise enough to
serve this role. \citet{wang2026landscape} identify two failure modes rooted in
prompt imprecision: \emph{mis-authorization}, where a user's explicit
instructions inadvertently grant the agent permission to follow injected
directives (e.g., ``read the file and follow its instructions''), and
\emph{semantic ambiguity}, where high-level prompts such as ``fix this bug''
provide insufficient specification for any downstream policy or alignment check
to reliably adjudicate edge cases. Performing alignment against an imprecise
anchor is building on sand, because errors in the ground truth propagate to every
decision the LLM judge makes.

\noindent\textbf{Reason 2: Some ambiguities are irreducible, and even a perfect model cannot resolve them.}
Certain security-relevant decisions hinge on user-specific context that is
simply absent from the prompt. Recent empirical work identifies two classes
of irreducible ambiguity. First, \emph{ambiguous language semantics}:
\citet{malaviya2025contextualized} show that the same query yields
different correct answers depending on context (user identity, expertise,
intent) that is not encoded in the prompt, and that this context can even
flip model rankings. Second, \emph{ambiguous objective alignment}: an
action that appears task-relevant may violate an implicit security boundary
the user never articulated. \citet{vijayvargiya2026ambig} demonstrate
that real-world agentic tasks contain multiple interdependent specification
gaps that agents cannot resolve without human
interaction---interactive clarification improves task success by up to
74\% over non-interactive settings. These ambiguities are not artifacts of
weak models; they arise from missing information that no model, however
capable, can manufacture~\citep{yang2025prompts}.

\noindent\textbf{Reason 3: LLM judges are themselves attackable, creating a circular dependency.}
Using an LLM to guard another LLM creates a circular
dependency~\citep{zhan2025adaptive}: if an attacker can manipulate the
agent LLM through prompt injection, the same vector can potentially compromise
the judge LLM. One mitigation is to tightly constrain both the inputs and
decision scope of the judge, but \citet{huang2025empirical} show
empirically that even fine-tuned judge models operate as task-specific
classifiers whose accuracy degrades sharply outside their training
distribution. Constrained LLM judges may thus be suitable for narrowly
scoped, structured artifacts, but not for open-ended judgment over arbitrary
text. The security-critical decisions that matter most are precisely those
that fall outside this narrow scope.

\noindent\textbf{Reason 4: Humans expressing intent directly is more reliable than LLM inference.}
When a user writes a policy (``only allow reading files in
\texttt{/project}''), approves an action, or configures a sandbox, they are
\emph{directly expressing} their intent rather than relying on an LLM to
\emph{infer} it from an ambiguous task description. Direct expression is
inherently more reliable: it eliminates the inference step entirely, removing
the LLM's uncertainty, its vulnerability to adversarial manipulation, and its
inability to access the user's private context.
Consider a concrete contrast. A user's scope rule \texttt{allow: read
/project/**} is deterministic and machine-verifiable; every file access can be
checked in constant time with zero ambiguity. An LLM judge that must decide
whether reading \texttt{/etc/passwd} is ``relevant to fixing the web app
bug'' faces an open-ended inference problem whose answer depends on unstated
assumptions about the user's deployment environment, threat model, and risk
tolerance. Each step in the inference chain introduces uncertainty, and
adversarial inputs can steer the chain toward the wrong conclusion.
Direct, user-specified policies (firewall rules, permission manifests,
access-control lists) have a long deployment history in traditional
security, and provide deterministic enforcement and explicit audit trails
by construction, properties that any inferred-intent mechanism must
approximate. Users do continue to engage meaningfully with such permission
and policy systems despite well-documented friction in their
design~\citep{prange2024not, tahaei2023stuck}; the widespread adoption
of policy specification, runtime approval, and scope configuration in
production agent systems is the latest instantiation of this pattern.

\noindent\textbf{Reason 5: Industry practice provides large-scale validation.}
The deployment data from \S\ref{sec:mismatch} constitute a natural experiment.
Twenty-one production systems, built independently by established AI
companies, startups, and open-source communities, have converged on
human-centric security architectures. Not a single
system has adopted fully automated intent anchoring (0/\nsystems).
This convergence is not coincidental; it reflects a practical reality: in
high-stakes, real-world deployment, human participation provides a reliability
floor that LLM-based approaches, in their current form, cannot match.

\subsection{But Current Human Participation Is Broken}
\label{sec:pos-antithesis}

If human participation is necessary, one might expect the problem to be
solved. It is not. Across all five \ahi categories, we observe a
fundamental \emph{cognitive burden vs.\ security guarantee} trade-off that
no existing mechanism resolves (\Cref{fig:burden-guarantee}).

\noindent\iconRuntimeApproval~\textbf{Runtime approval (high guarantee, high burden).}
Runtime approval is the most widely deployed mechanism (15/\nsystems) and
provides strong security guarantees, since every sensitive action is gated on
explicit human consent. But it imposes the highest cognitive burden: users
become ``OK-button machines.'' In a typical agentic coding session with
\claudecode, a single task such as ``refactor the authentication module'' can
trigger dozens of approval prompts---one for each file write, shell command,
and API call---forcing the developer to context-switch repeatedly between
reviewing agent actions and their own work. This mirrors the well-documented
phenomenon of \emph{warning fatigue} in browser security, where users
click through 90\%+ of SSL warnings without reading
them~\citep{alahmadi202299}. Every fatigue-mitigation strategy
deployed today has drawbacks: ``always allow'' silently expands the trusted
surface~\cite{anthropic2025claudecode}; sandbox auto-approval sacrifices
flexibility; and risk-adaptive screening~\cite{zhong2025rtbas}, the only
principled alternative, remains academic with zero production adoption.

\noindent\iconScopeCfg~\textbf{Scope configuration (moderate both, but drifts).}
Scope configuration (16/\nsystems) offers a moderate position, but suffers
from \emph{scope creep}: ``always allow'' clicks and session-wide
authorizations monotonically widen boundaries with no principled
shrink-back mechanism. The consequences can be severe: Replit Agent's
production database deletion incident (July 2025) was traced to an
over-broad sandbox configuration that granted write access to production
resources during a development task~\cite{replit2025agent}. The
default-stance dilemma further splits the design space: default-open
(e.g., \aider{}, which provides no sandbox at all~\cite{aider2025}) is
frictionless but incident-prone; default-closed (e.g., \codex's
network-isolated container~\cite{openai2025codex}) is safe but
obstructive~\cite{anthropic2025claudecodesandbox}. No production system
today offers a principled middle ground with bidirectional scope
adjustment.

\noindent\iconPolicySpec~\textbf{Policy specification (moderate--high guarantee, high burden for users).}
Policy specification (14/\nsystems) can provide strong guarantees when
policies are formal~\cite{shi2025progent}, but formal DSLs are
inaccessible to ordinary users. Even state-of-the-art automated policy
generation falls short: AgentSpec's LLM-based rule generation achieves only
70.96\% recall~\cite{wang2026agentspec}, meaning nearly 30\% of necessary
constraints are missed entirely. Production systems therefore default to
natural-language policies (\claudecode's \texttt{CLAUDE.md}, \cursorsys's
\texttt{.cursorrules}) whose enforcement reduces to LLM
self-compliance---the agent interprets the policy using the same LLM
capabilities that may be compromised by prompt injection, which is no
better than asking the agent to promise to behave.

\noindent\iconTrustLabel~\textbf{Trust labeling (high burden, zero adoption).}
Trust labeling imposes the highest conceptual burden: users must reason about
data provenance at a granularity (trust lattices, information-flow labels)
that mismatches their mental models. Zero of \nsystems production systems
expose trust labeling as a user-facing feature, even though low-burden
variants exist~\cite{siddiqui2024permissive, debenedetti2025defeating}.
The barrier is not effort alone; it is \emph{conceptual mismatch}.

\noindent\iconIntentAnchor~\textbf{Intent anchoring (zero burden, zero adoption).}
Intent anchoring requires nothing from the user, as security enforcement is
invisible. Paradoxically, this invisibility may \emph{impede} adoption:
users cannot perceive the safeguard, cannot calibrate their trust in it, and
cannot diagnose false positives. A system that blocks a legitimate action
without explanation (because the user never asked for a security check) erodes
trust rather than building it.

\medskip\noindent
The pattern is stark: \textbf{no existing \ahi category simultaneously
achieves low cognitive burden and high security guarantee}. This is not a
design flaw of individual mechanisms; it is the consequence of a field that
has not systematically studied the human side of agent security.

\subsection{Synthesis: \ahi{} Security as a First-Class Citizen}
\label{sec:pos-synthesis}

The preceding analysis yields a clear logical chain:

\begin{enumerate}[noitemsep, topsep=1pt, leftmargin=*]
  \item Human participation in agent security decisions is \textbf{necessary}
  (\S\ref{sec:pos-thesis}): LLM-based intent alignment alone faces
  irreducible limitations (imprecise anchors, fundamental ambiguity,
  circular vulnerability), direct expression of intent is a more reliable
  substrate than LLM inference, and twenty-one production deployments
  independently validate this conclusion.
  \item Current human participation is \textbf{broken}
  (\S\ref{sec:pos-antithesis}): every deployed mechanism forces users into an
  unsatisfactory trade-off between cognitive burden and security guarantee.
  \item Therefore, the field needs \textbf{systematic research} into
  designing effective, low-burden human participation mechanisms for agent
  security.
\end{enumerate}

This research agenda is neither a pure HCI problem nor a pure security
problem. HCI alone cannot address it, because designing effective agent
security interactions requires understanding threat models, access-control
semantics, and the adversarial dynamics of prompt injection. Security alone
cannot address it, because building usable mechanisms requires understanding
cognitive load, mental models, and human decision-making under uncertainty.
\ahisec is an inherently \textbf{interdisciplinary} challenge that demands its
own design principles, evaluation methods, and theoretical foundations.

We call on the AI community to elevate \ahisec from an implementation
afterthought to a \textbf{first-class research citizen}, a recognized
subfield at the intersection of security, human--computer interaction, and
artificial intelligence.

\section{Alternative Views}
\label{sec:alt-views}

We anticipate three principal objections to our position and address each in
turn.

\noindent\textbf{Objection 1: ``LLM capabilities are rapidly improving; full automation will be possible soon.''}
It is true that model capabilities are advancing quickly, and future LLMs will
likely reduce the \emph{rate} of alignment errors. However, two of our five
reasons are capability-independent. Irreducible ambiguity
(\S\ref{sec:pos-thesis}, Reason~2) arises from missing user context, not
from model limitations: empirical studies confirm that underspecified prompts
cause failures even in the strongest models~\citep{yang2025prompts}, and
that user-specific context (identity, expertise, intent) can flip which
response is correct~\citep{malaviya2025contextualized}---no amount of
capability closes this gap without human input. Circular dependency (Reason~3) is
structural, since using an LLM to guard an LLM does not become safe simply
because both LLMs grow more capable; the attack surface scales with
capability. Improved models will make Path~A more useful as a
\emph{complement} to human participation, but they will not eliminate the
need for it.

\noindent\textbf{Objection 2: ``Human-in-the-loop cannot scale.''}
This objection correctly identifies the central challenge of
\S\ref{sec:pos-antithesis}: current human participation mechanisms impose
unsustainable cognitive burden. But the appropriate response is to make human
participation more efficient, not to remove humans from the loop. The
usable-security community faced an analogous dilemma with SSL certificate
warnings in the 2000s: early warnings were ignored by the vast majority of
users, leading some to argue that browser security warnings should be
eliminated. Instead, researchers redesigned the interaction by simplifying
language, reducing warning frequency, and making safe defaults
automatic, and dramatically improved both compliance and
security~\cite{felt2015improving}. Our research agenda
(\S\ref{sec:agenda}) follows the same philosophy: the solution to bad human
participation is better human participation, not no human participation.

\noindent\textbf{Objection 3: ``This is just an HCI problem; it does not belong in AI security venues.''}
We argue the opposite: \ahisec is precisely the kind of interdisciplinary
problem that AI security venues are best positioned to address. Pure HCI
research can study cognitive load and interaction design, but it lacks the
domain expertise to model agent threat landscapes, formalize access-control
semantics, or reason about adversarial prompt injection. Pure security
research can formalize threat models, but it routinely assumes a
``perfect user'' who reads every warning, understands every permission dialog,
and never clicks ``allow'' out of fatigue. Neither community can solve this
problem alone. \ahisec requires researchers who understand \emph{both} the
security semantics of agent actions \emph{and} the cognitive constraints of
the humans who must oversee them. Venues like NeurIPS, which bridge AI,
security, and human factors, are the natural home for this work.

\section{A Human-Centric Research Agenda}
\label{sec:agenda}

The cognitive burden vs.\ security guarantee trade-off identified in
\S\ref{sec:pos-antithesis} is not inevitable; it is the product of a field
that has not yet invested systematic effort in the human side of agent
security. We propose three research directions, each targeting a specific
gap in the trade-off space.

\subsection{Direction 1: Fatigue-Resistant \ahi{} Mechanisms}
\label{sec:agenda-fatigue}

\textbf{Motivated by:} the finding that runtime approval imposes the highest
cognitive burden (approval fatigue) and scope configuration suffers from expand-only
drift (scope creep), yet both are the most widely deployed categories
(15 and 16 of \nsystems systems, respectively).
The Replit production-database deletion incident (July 2025)~\citep{replit2025agent} is a concrete instantiation of how cumulative scope expansion can turn a routine development task into a destructive event.

\textbf{Core question:} How can we reduce the human decision burden without
sacrificing security guarantees?
Today's mechanisms force a Pareto trade-off between high friction with strong guarantees (default-prompt) and low friction with weak guarantees (always-allow), so progress requires shifting that frontier rather than choosing a point along it.

Two sub-directions are especially promising.
\emph{Risk-adaptive approval} dynamically calibrates whether an action requires
human confirmation based on its assessed risk level: low-risk operations (e.g.,
reading a file already in the project directory) are auto-approved, while
high-risk operations (e.g., executing a shell command that modifies system
files) trigger an explicit approval prompt. RTBAS~\cite{zhong2025rtbas}
demonstrates the feasibility of this approach, preventing all targeted attacks
with only 2\% utility loss in controlled experiments, yet no production system
has adopted it. Understanding and closing this adoption gap is a concrete
research opportunity.
\emph{Bidirectional scope adjustment} addresses the expand-only drift problem by
giving users explicit mechanisms to \emph{shrink} scope, such as revoking
over-authorizations, tightening sandbox boundaries mid-session, and receiving
notifications when cumulative ``always allow'' decisions have materially
expanded the attack surface. Today, only two systems (\cursorsys{} admin policy and
OpenClaw) support any form of scope contraction.

\subsection{Direction 2: Usable Policy and Trust Models}
\label{sec:agenda-usable}

\textbf{Motivated by:} the finding that policy specification forces a choice between
unreliable natural-language policies and unusable formal DSLs, and that
trust labeling has zero adoption due to conceptual mismatch with users' mental
models.
Both gaps share a common root: the field has assumed users will adapt to the system's abstractions rather than designing abstractions that meet users where they are.

\textbf{Core question:} How can we bridge the gap between policies that
users can write and policies that systems can enforce?
The question splits along authoring (what natural-language input can be reliably compiled?) and enforcement (how should the compiled policy be validated by its author?) axes that must be tackled jointly.

\emph{NL-to-formal policy translation} offers a promising path: users
author policies in natural language (as they already do with
\texttt{CLAUDE.md}), and a translation layer compiles them into
enforceable formal constraints. This combines the authoring ease of natural
language with the enforcement reliability of deterministic engines, though it
introduces new challenges around translation fidelity and user verification
of compiled policies.
\emph{Progressive trust models} address the trust labeling gap by meeting users
where they are. Rather than requiring users to reason about lattice-based
information-flow control, progressive models start with simple binary
distinctions (trusted vs.\ untrusted sources) and incrementally introduce
finer-grained labels as users develop expertise. The goal is to align the
system's trust abstraction with users' actual mental models of data
provenance.

\subsection{Direction 3: LLM-Augmented Human Intent Alignment}
\label{sec:agenda-augmented}

\textbf{Motivated by:} the synthesis in \S\ref{sec:pos-synthesis}, that LLMs
should augment human intent alignment, not replace it.
The three failure modes from \S\ref{sec:pos-thesis} (prompt imprecision, irreducible ambiguity, circular dependency) rule out LLMs as final authorities on intent, but they do not rule out LLMs as accelerators of human judgment.

\textbf{Core question:} How can LLMs help humans express and confirm
intent more efficiently, rather than attempting to judge intent autonomously?
This reframes the design problem from ``can the LLM be trusted to decide?'' to ``where in the human's intent-alignment workflow can the LLM contribute the most leverage at the lowest risk?''

\S\ref{sec:pos-thesis} argued that LLMs alone cannot reliably perform intent
alignment. But this does not mean LLMs have no role; it means their role
should be \emph{assistive} rather than \emph{decisional}.
\emph{LLM-generated policy drafts} exemplify this reframing: rather than
writing policies from scratch, users review and edit LLM-generated drafts
tailored to their task. AutoPermissions~\cite{wu2025towards}
demonstrates that LLMs can predict user approval decisions with 85.1\%
accuracy from context, suggesting that LLM-drafted policies could
substantially reduce authoring burden while keeping the human as the final
authority.
\emph{Proactive ambiguity detection} inverts the typical interaction pattern:
instead of waiting for a security-relevant decision to arise at runtime and
then interrupting the user, the system analyzes the task description
\emph{before} execution, identifies points of ambiguity or potential
over-authorization, and front-loads human clarification into the planning
phase. This reduces runtime interruptions, the primary driver of approval
fatigue, by resolving ambiguities when the user is already engaged in
task specification rather than mid-execution.

\medskip
\noindent
Extended sub-directions for all three research areas, including cognitive
load-aware interaction design, template-based policy authoring, preference
learning from approval histories, and structured intent elicitation, are
detailed in Appendix~\ref{app:agenda-extended}.

\medskip
\noindent
The three directions are complementary rather than independent.
Fatigue-resistant mechanisms (Direction~1) reduce the volume of decisions a user must make, but each remaining decision still depends on a usable abstraction in which to express it---the province of Direction~2.
LLM-augmented drafting (Direction~3) further reduces authoring burden across both, by turning policy specification, scope configuration, and approval into \emph{edit-rather-than-write} tasks.
Pursued together, they describe a single research program: keep the human as the final authority on intent, but design the surrounding interaction so that authority is exercised at the right moments, on the right abstractions, with the right assistance.

\section{Limitations}
\label{sec:limitations}

Our work has several limitations.
First, the SoK corpus is bounded by our search scope (top-tier security, systems, ML, SE, and NLP venues plus arXiv through \datayear) and may miss work in other venues or industry grey literature.
Second, our analysis of production systems relies on public documentation, engineering blogs, and system cards; undisclosed internals may include additional mechanisms.
Third, while our five-category \ahi{} framework covers all mechanisms we encountered, the taxonomy is not guaranteed to remain exhaustive as the field evolves.
Finally, our research agenda is conceptual: we motivate directions from empirical observations but have not yet conducted user studies or built prototypes.

\section{Conclusion}
\label{sec:conclusion}

The core challenge of LLM agent security is intent alignment, and our analysis shows it cannot be solved without human participation: LLM-based alignment faces irreducible limitations (prompt imprecision, fundamental ambiguity, circular vulnerability), while humans remain the only available ground-truth oracle for their own intent. Industry has already converged on this conclusion. Across \nsystems production systems, three human-centric mechanisms (policy specification, runtime approval, scope configuration) form the backbone of deployed agent security, while the two academia-favored categories (intent anchoring and trust labeling) see zero production adoption---a pronounced and consequential industry--academia mismatch.

Yet current human participation is far from satisfactory: every deployed mechanism forces users into an unsatisfying trade-off between cognitive burden and security guarantee. The path forward is not to remove humans from the loop but to make their participation more effective, efficient, and informed---through fatigue-resistant interaction, usable policy and trust models, and LLM-augmented intent alignment. We call on the AI community to elevate \ahisec{} to a first-class research citizen.

\begin{ack}
\end{ack}

\newpage
\clearpage

{\small
\bibliographystyle{plainnat}
\bibliography{main}

\begin{thebibliography}{68}
\providecommand{\natexlab}[1]{#1}
\providecommand{\url}[1]{\texttt{#1}}
\expandafter\ifx\csname urlstyle\endcsname\relax
  \providecommand{\doi}[1]{doi: #1}\else
  \providecommand{\doi}{doi: \begingroup \urlstyle{rm}\Url}\fi

\bibitem[Abaev et~al.(2026)Abaev, Klimov, Levinov, Mimran, Elovici, and
  Shabtai]{abaev2026agentguardian}
Nadya Abaev, Denis Klimov, Gerard Levinov, David Mimran, Yuval Elovici, and
  Asaf Shabtai.
\newblock Agentguardian: Learning access control policies to govern ai agent
  behavior.
\newblock \emph{arXiv preprint arXiv:2601.10440}, 2026.

\bibitem[{Adversa AI}(2025)]{secureclaw2025}
{Adversa AI}.
\newblock {SecureClaw}: {OWASP}-aligned security plugin.
\newblock \url{https://github.com/adversa-ai/secureclaw}, 2025.

\bibitem[{Aider}(2025)]{aider2025}
{Aider}.
\newblock Aider: Terminal-based {AI} pair programming.
\newblock \url{https://aider.chat}, 2025.

\bibitem[Alahmadi et~al.(2022)Alahmadi, Axon, and Martinovic]{alahmadi202299}
Bushra~A Alahmadi, Louise Axon, and Ivan Martinovic.
\newblock 99\% false positives: A qualitative study of $\{$SOC$\}$ analysts'
  perspectives on security alarms.
\newblock In \emph{31st USENIX Security Symposium (USENIX Security 22)}, pages
  2783--2800, 2022.

\bibitem[{Amazon Web Services}(2025)]{aws2025bedrock}
{Amazon Web Services}.
\newblock Amazon bedrock agents.
\newblock \url{https://aws.amazon.com/bedrock/agents/}, 2025.

\bibitem[{Anthropic}(2025{\natexlab{a}})]{anthropic2025claudecode}
{Anthropic}.
\newblock Claude code: Anthropic's agentic coding tool.
\newblock \url{https://docs.anthropic.com/en/docs/claude-code},
  2025{\natexlab{a}}.

\bibitem[{Anthropic}(2025{\natexlab{b}})]{anthropic2025claudecodesandbox}
{Anthropic}.
\newblock Claude code sandboxing.
\newblock \url{https://www.anthropic.com/engineering/claude-code-sandboxing},
  2025{\natexlab{b}}.

\bibitem[B{\"u}hler et~al.(2025)B{\"u}hler, Biagiola, Di~Grazia, and
  Salvaneschi]{buhler2025securing}
Christoph B{\"u}hler, Matteo Biagiola, Luca Di~Grazia, and Guido Salvaneschi.
\newblock Securing ai agent execution.
\newblock \emph{arXiv preprint arXiv:2510.21236}, 2025.

\bibitem[Christodorescu et~al.(2025)Christodorescu, Fernandes, Hooda, Jha,
  Rehberger, Chaudhuri, Fu, Shams, Amir, Choi,
  et~al.]{christodorescu2025systems}
Mihai Christodorescu, Earlence Fernandes, Ashish Hooda, Somesh Jha, Johann
  Rehberger, Kamalika Chaudhuri, Xiaohan Fu, Khawaja Shams, Guy Amir, Jihye
  Choi, et~al.
\newblock Systems security foundations for agentic computing.
\newblock \emph{arXiv preprint arXiv:2512.01295}, 2025.

\bibitem[{Cognition}(2025)]{devin2025cognition}
{Cognition}.
\newblock Devin: Autonomous {AI} software engineer.
\newblock \url{https://devin.ai/}, 2025.

\bibitem[Costa et~al.(2025)Costa, K{\"o}pf, Kolluri, Paverd, Russinovich,
  Salem, Tople, Wutschitz, and Zanella-B{\'e}guelin]{costa2025securing}
Manuel Costa, Boris K{\"o}pf, Aashish Kolluri, Andrew Paverd, Mark Russinovich,
  Ahmed Salem, Shruti Tople, Lukas Wutschitz, and Santiago
  Zanella-B{\'e}guelin.
\newblock Securing ai agents with information-flow control.
\newblock \emph{arXiv preprint arXiv:2505.23643}, 2025.

\bibitem[{CrewAI}(2025)]{crewai2025}
{CrewAI}.
\newblock {CrewAI}: Multi-agent orchestration framework.
\newblock \url{https://crewai.com/}, 2025.

\bibitem[{Cursor}(2025)]{cursor2025security}
{Cursor}.
\newblock Cursor {AI}-native {IDE} security.
\newblock \url{https://cursor.com/security}, 2025.

\bibitem[Debenedetti et~al.(2025)Debenedetti, Shumailov, Fan, Hayes, Carlini,
  Fabian, Kern, Shi, Terzis, and Tram{\`e}r]{debenedetti2025defeating}
Edoardo Debenedetti, Ilia Shumailov, Tianqi Fan, Jamie Hayes, Nicholas Carlini,
  Daniel Fabian, Christoph Kern, Chongyang Shi, Andreas Terzis, and Florian
  Tram{\`e}r.
\newblock Defeating prompt injections by design.
\newblock \emph{arXiv preprint arXiv:2503.18813}, 2025.

\bibitem[Doshi et~al.(2026)Doshi, Hong, Xu, Kang, Kapravelos, and
  K{\"a}stner]{doshi2026towards}
Aarya Doshi, Yining Hong, Congying Xu, Eunsuk Kang, Alexandros Kapravelos, and
  Christian K{\"a}stner.
\newblock Towards verifiably safe tool use for llm agents.
\newblock \emph{arXiv preprint arXiv:2601.08012}, 2026.

\bibitem[{Eqty Lab}(2025)]{mcpguardian2025}
{Eqty Lab}.
\newblock mcp-guardian: {MCP} proxy with human approval.
\newblock \url{https://github.com/eqtylab/mcp-guardian}, 2025.

\bibitem[Felt et~al.(2015)Felt, Ainslie, Reeder, Consolvo, Thyagaraja, Bettes,
  Harris, and Grimes]{felt2015improving}
Adrienne~Porter Felt, Alex Ainslie, Robert~W Reeder, Sunny Consolvo, Somas
  Thyagaraja, Alan Bettes, Helen Harris, and Jeff Grimes.
\newblock Improving ssl warnings: Comprehension and adherence.
\newblock In \emph{Proceedings of the 33rd annual ACM conference on human
  factors in computing systems}, pages 2893--2902, 2015.

\bibitem[Feng et~al.(2025)Feng, Kim, Pang, Huq, August, and
  Zhang]{feng2025regulatory}
KJ~Feng, Tae~Soo Kim, Rock~Yuren Pang, Faria Huq, Tal August, and Amy~X Zhang.
\newblock On the regulatory potential of user interfaces for ai agent
  governance.
\newblock \emph{arXiv preprint arXiv:2512.00742}, 2025.

\bibitem[{GitHub}(2025)]{github2025copilotcli}
{GitHub}.
\newblock Github copilot {CLI}.
\newblock \url{https://github.com/features/copilot/cli}, 2025.

\bibitem[Gong et~al.(2025)Gong, Li, Chang, and Shen]{gong2025csagent}
Haochen Gong, Chenxiao Li, Rui Chang, and Wenbo Shen.
\newblock Secure and efficient access control for computer-use agents via
  context space.
\newblock \emph{arXiv preprint arXiv:2509.22256}, 2025.

\bibitem[{Google}(2025{\natexlab{a}})]{google2025adk}
{Google}.
\newblock Google agent development kit ({ADK}).
\newblock \url{https://github.com/google/adk-python}, 2025{\natexlab{a}}.

\bibitem[{Google}(2025{\natexlab{b}})]{google2025geminicli}
{Google}.
\newblock Gemini {CLI}.
\newblock \url{https://gemini.google.com/}, 2025{\natexlab{b}}.

\bibitem[{Guardrails AI}(2025)]{guardrailsai2025}
{Guardrails AI}.
\newblock Guardrails {AI}: {Python} framework for {LLM} validation.
\newblock \url{https://github.com/guardrails-ai/guardrails}, 2025.

\bibitem[Huang et~al.(2025)Huang, Bu, Zhou, Qu, Liu, Yang, Xu, and
  Zhao]{huang2025empirical}
Hui Huang, Xingyuan Bu, Hongli Zhou, Yingqi Qu, Jing Liu, Muyun Yang, Bing Xu,
  and Tiejun Zhao.
\newblock An empirical study of llm-as-a-judge for llm evaluation: Fine-tuned
  judge model is not a general substitute for gpt-4.
\newblock In \emph{Findings of the Association for Computational Linguistics:
  ACL 2025}, pages 5880--5895, 2025.

\bibitem[Huang et~al.(2024)Huang, Liu, Chen, Wang, Wang, Lian, Wang, Tang, and
  Chen]{huang2024understanding}
Xu~Huang, Weiwen Liu, Xiaolong Chen, Xingmei Wang, Hao Wang, Defu Lian, Yasheng
  Wang, Ruiming Tang, and Enhong Chen.
\newblock Understanding the planning of llm agents: A survey.
\newblock \emph{arXiv preprint arXiv:2402.02716}, 2024.

\bibitem[{Invariant Labs / Snyk}(2025)]{mcpscan2025}
{Invariant Labs / Snyk}.
\newblock mcp-scan: {MCP} server security scanner.
\newblock \url{https://github.com/invariantlabs-ai/mcp-scan}, 2025.

\bibitem[Ji et~al.(2026)Ji, Wu, Jiang, Ma, Li, Gao, Wang, and Li]{ji2026taming}
Zimo Ji, Daoyuan Wu, Wenyuan Jiang, Pingchuan Ma, Zongjie Li, Yudong Gao, Shuai
  Wang, and Yingjiu Li.
\newblock Taming various privilege escalation in llm-based agent systems: A
  mandatory access control framework.
\newblock \emph{arXiv preprint arXiv:2601.11893}, 2026.

\bibitem[Jia et~al.(2025)Jia, Wu, Qin, and Squicciarini]{jia2025task}
Feiran Jia, Tong Wu, Xin Qin, and Anna Squicciarini.
\newblock The task shield: Enforcing task alignment to defend against indirect
  prompt injection in llm agents.
\newblock In \emph{Proceedings of the 63rd Annual Meeting of the Association
  for Computational Linguistics (Volume 1: Long Papers)}, pages 29680--29697,
  2025.

\bibitem[kenryu42(2025)]{safetynet2025}
kenryu42.
\newblock claude-code-safety-net: Hook for blocking dangerous commands.
\newblock \url{https://github.com/kenryu42/claude-code-safety-net}, 2025.

\bibitem[Kim et~al.(2025)Kim, Choi, and Lee]{kim2025prompt}
Juhee Kim, Woohyuk Choi, and Byoungyoung Lee.
\newblock Prompt flow integrity to prevent privilege escalation in llm agents.
\newblock \emph{arXiv preprint arXiv:2503.15547}, 2025.

\bibitem[{LangChain}(2025)]{langchain2025}
{LangChain}.
\newblock {LangChain}/{LangGraph} agent framework.
\newblock \url{https://www.langchain.com/langgraph}, 2025.

\bibitem[Li et~al.(2025{\natexlab{a}})Li, Mallick, Rose, Robertson, Oprea, and
  Nita-Rotaru]{li2025ace}
Evan Li, Tushin Mallick, Evan Rose, William Robertson, Alina Oprea, and
  Cristina Nita-Rotaru.
\newblock Ace: A security architecture for llm-integrated app systems.
\newblock \emph{arXiv preprint arXiv:2504.20984}, 2025{\natexlab{a}}.

\bibitem[Li et~al.(2025{\natexlab{b}})Li, Liu, Chiu, Li, Zhang, and
  Xiao]{li2025drift}
Hao Li, Xiaogeng Liu, Hung-Chun Chiu, Dianqi Li, Ning Zhang, and Chaowei Xiao.
\newblock Drift: Dynamic rule-based defense with injection isolation for
  securing llm agents.
\newblock \emph{arXiv preprint arXiv:2506.12104}, 2025{\natexlab{b}}.

\bibitem[Li et~al.(2025{\natexlab{c}})Li, Zou, Wu, Li, Xing, Zheng, Hu, Wang,
  Li, Yuan, et~al.]{li2025safeflow}
Peiran Li, Xinkai Zou, Zhuohang Wu, Ruifeng Li, Shuo Xing, Hanwen Zheng, Zhikai
  Hu, Yuping Wang, Haoxi Li, Qin Yuan, et~al.
\newblock Safeflow: A principled protocol for trustworthy and transactional
  autonomous agent systems.
\newblock \emph{arXiv preprint arXiv:2506.07564}, 2025{\natexlab{c}}.

\bibitem[Luo et~al.(2025)Luo, Dai, Liu, Banerjee, Sun, Chen, and
  Xiao]{luo2025agrail}
Weidi Luo, Shenghong Dai, Xiaogeng Liu, Suman Banerjee, Huan Sun, Muhao Chen,
  and Chaowei Xiao.
\newblock Agrail: A lifelong agent guardrail with effective and adaptive safety
  detection.
\newblock In \emph{Proceedings of the 63rd Annual Meeting of the Association
  for Computational Linguistics (Volume 1: Long Papers)}, pages 8104--8139,
  2025.

\bibitem[Malaviya et~al.(2025)Malaviya, Chang, Roth, Iyyer, Yatskar, and
  Lo]{malaviya2025contextualized}
Chaitanya Malaviya, Joseph~Chee Chang, Dan Roth, Mohit Iyyer, Mark Yatskar, and
  Kyle Lo.
\newblock Contextualized evaluations: Judging language model responses to
  underspecified queries.
\newblock \emph{Transactions of the Association for Computational Linguistics},
  13:\penalty0 878--900, 2025.

\bibitem[{Microsoft}(2025)]{microsoft2025agentfw}
{Microsoft}.
\newblock Microsoft agent framework autogen.
\newblock \url{https://github.com/microsoft/autogen}, 2025.

\bibitem[{OpenAI}(2025)]{openai2025codex}
{OpenAI}.
\newblock Codex: {OpenAI}'s cloud coding agent.
\newblock \url{https://developers.openai.com/codex/agent-approvals-security},
  2025.

\bibitem[Pawelek et~al.(2025)Pawelek, Patel, Crowell, Golilarz, Mittal, Rahimi,
  and Perkins]{pawelek2025llmzplus}
Tom Pawelek, Raj Patel, Charlotte Crowell, Noorbakhsh~Amiri Golilarz, Sudip
  Mittal, Shahram Rahimi, and Andy Perkins.
\newblock Llmz+: Contextual prompt whitelist principles for agentic llms.
\newblock In \emph{2025 International Conference on Machine Learning and
  Applications (ICMLA)}, pages 1396--1402. IEEE, 2025.

\bibitem[{PolicyLayer}(2025)]{intercept2025}
{PolicyLayer}.
\newblock Intercept: {YAML} policy enforcement for {MCP}.
\newblock \url{https://github.com/PolicyLayer/Intercept}, 2025.

\bibitem[Prange et~al.(2024)Prange, Knierim, Knoll, Dietz, De~Luca, and
  Alt]{prange2024not}
Sarah Prange, Pascal Knierim, Gabriel Knoll, Felix Dietz, Alexander De~Luca,
  and Florian Alt.
\newblock $\{$“I$\}$ do (not) need that $\{$Feature!”$\}$--understanding
  $\{$Users’$\}$ awareness and control of privacy permissions on android
  smartphones.
\newblock In \emph{Twentieth Symposium on Usable Privacy and Security (SOUPS
  2024)}, pages 453--472, 2024.

\bibitem[Rebedea et~al.(2023)Rebedea, Dinu, Sreedhar, Parisien, and
  Cohen]{rebedea2023nemo}
Traian Rebedea, Razvan Dinu, Makesh~Narsimhan Sreedhar, Christopher Parisien,
  and Jonathan Cohen.
\newblock Nemo guardrails: A toolkit for controllable and safe llm applications
  with programmable rails.
\newblock In \emph{Proceedings of the 2023 conference on empirical methods in
  natural language processing: system demonstrations}, pages 431--445, 2023.

\bibitem[{Replit}(2025)]{replit2025agent}
{Replit}.
\newblock Replit agent.
\newblock \url{https://blog.replit.com/safe-vibe-coding}, 2025.

\bibitem[{Salesforce}(2025)]{salesforce2025agentforce}
{Salesforce}.
\newblock Agentforce: {Salesforce} enterprise agent platform.
\newblock
  \url{https://trailhead.salesforce.com/content/learn/modules/trusted-agentic-ai/explore-agentforce-guardrails-and-trust-patterns},
  2025.

\bibitem[Sequeira et~al.(2026)Sequeira, Damianakis, Iqbal, and
  Psounis]{sequeira2026agent}
Rohan Sequeira, Stavros Damianakis, Umar Iqbal, and Konstantinos Psounis.
\newblock Agent-sentry: Bounding llm agents via execution provenance.
\newblock \emph{arXiv preprint arXiv:2603.22868}, 2026.

\bibitem[Shan et~al.(2026)Shan, Xin, Zhang, and Xu]{shan2026don}
Zhengyang Shan, Jiayun Xin, Yue Zhang, and Minghui Xu.
\newblock Don't let the claw grip your hand: A security analysis and defense
  framework for openclaw.
\newblock \emph{arXiv preprint arXiv:2603.10387}, 2026.

\bibitem[Shi et~al.(2025)Shi, He, Wang, Li, Wu, Guo, and Song]{shi2025progent}
Tianneng Shi, Jingxuan He, Zhun Wang, Hongwei Li, Linyu Wu, Wenbo Guo, and Dawn
  Song.
\newblock Progent: Programmable privilege control for llm agents.
\newblock \emph{arXiv preprint arXiv:2504.11703}, 2025.

\bibitem[Siddiqui et~al.(2024)Siddiqui, Gaonkar, K{\"o}pf, Krueger, Paverd,
  Salem, Tople, Wutschitz, Xia, and
  Zanella-B{\'e}guelin]{siddiqui2024permissive}
Shoaib~Ahmed Siddiqui, Radhika Gaonkar, Boris K{\"o}pf, David Krueger, Andrew
  Paverd, Ahmed Salem, Shruti Tople, Lukas Wutschitz, Menglin Xia, and Santiago
  Zanella-B{\'e}guelin.
\newblock Permissive information-flow analysis for large language models.
\newblock \emph{arXiv preprint arXiv:2410.03055}, 2024.

\bibitem[Tahaei et~al.(2023)Tahaei, Abu-Salma, and Rashid]{tahaei2023stuck}
Mohammad Tahaei, Ruba Abu-Salma, and Awais Rashid.
\newblock Stuck in the permissions with you: Developer \& end-user perspectives
  on app permissions \& their privacy ramifications.
\newblock In \emph{Proceedings of the 2023 CHI Conference on Human Factors in
  Computing Systems}, pages 1--24, 2023.

\bibitem[{Trail of Bits}(2025)]{trailofbits2025skills}
{Trail of Bits}.
\newblock Trail of bits security skills for claude code.
\newblock \url{https://github.com/trailofbits/skills}, 2025.

\bibitem[Tsai and Bagdasarian(2025)]{tsai2025contextual}
Lillian Tsai and Eugene Bagdasarian.
\newblock Contextual agent security: A policy for every purpose.
\newblock In \emph{Proceedings of the 2025 Workshop on Hot Topics in Operating
  Systems}, pages 8--17, 2025.

\bibitem[Vijayvargiya et~al.(2026)Vijayvargiya, Zhou, Yerukola, Sap, and
  Neubig]{vijayvargiya2026ambig}
Sanidhya Vijayvargiya, Xuhui Zhou, Akhila Yerukola, Maarten Sap, and Graham
  Neubig.
\newblock Ambig-swe: Interactive agents to overcome underspecificity in
  software engineering.
\newblock In \emph{The Fourteenth International Conference on Learning
  Representations}, 2026.

\bibitem[Wang et~al.(2026{\natexlab{a}})Wang, Poskitt, and
  Sun]{wang2026agentspec}
Haoyu Wang, Christopher~M Poskitt, and Jun Sun.
\newblock Agentspec: Customizable runtime enforcement for safe and reliable llm
  agents.(2026).
\newblock In \emph{Proceedings of the IEEE/ACM International Conference on
  Software Engineering, ICSE}, pages 12--18, 2026{\natexlab{a}}.

\bibitem[Wang et~al.(2024)Wang, Wu, Li, Pan, Suh, Mao, Chen, and
  Xiao]{wang2024fath}
Jiongxiao Wang, Fangzhou Wu, Wendi Li, Jinsheng Pan, Edward Suh, Z~Morley Mao,
  Muhao Chen, and Chaowei Xiao.
\newblock Fath: Authentication-based test-time defense against indirect prompt
  injection attacks.
\newblock \emph{arXiv preprint arXiv:2410.21492}, 2024.

\bibitem[Wang et~al.(2025)Wang, Liu, Lu, Cai, Chen, Yang, Zhang, Hong, and
  Wu]{wang2025agentarmor}
Peiran Wang, Yang Liu, Yunfei Lu, Yifeng Cai, Hongbo Chen, Qingyou Yang, Jie
  Zhang, Jue Hong, and Ye~Wu.
\newblock Agentarmor: Enforcing program analysis on agent runtime trace to
  defend against prompt injection.
\newblock \emph{arXiv preprint arXiv:2508.01249}, 2025.

\bibitem[Wang et~al.(2026{\natexlab{b}})Wang, Li, Xiang, Zhang, Li, Zhang,
  Wang, and Tian]{wang2026landscape}
Peiran Wang, Xinfeng Li, Chong Xiang, Jinghuai Zhang, Ying Li, Lixia Zhang,
  Xiaofeng Wang, and Yuan Tian.
\newblock The landscape of prompt injection threats in llm agents: From
  taxonomy to analysis.
\newblock \emph{arXiv preprint arXiv:2602.10453}, 2026{\natexlab{b}}.

\bibitem[Wu et~al.(2024{\natexlab{a}})Wu, Cecchetti, and Xiao]{wu2024system}
Fangzhou Wu, Ethan Cecchetti, and Chaowei Xiao.
\newblock System-level defense against indirect prompt injection attacks: An
  information flow control perspective.
\newblock \emph{arXiv preprint arXiv:2409.19091}, 2024{\natexlab{a}}.

\bibitem[Wu et~al.(2024{\natexlab{b}})Wu, Roesner, Kohno, Zhang, and
  Iqbal]{wu2024isolategpt}
Yuhao Wu, Franziska Roesner, Tadayoshi Kohno, Ning Zhang, and Umar Iqbal.
\newblock Isolategpt: An execution isolation architecture for llm-based agentic
  systems.
\newblock \emph{arXiv preprint arXiv:2403.04960}, 2024{\natexlab{b}}.

\bibitem[Wu et~al.(2025)Wu, Yang, Roesner, Kohno, Zhang, and
  Iqbal]{wu2025towards}
Yuhao Wu, Ke~Yang, Franziska Roesner, Tadayoshi Kohno, Ning Zhang, and Umar
  Iqbal.
\newblock Towards automating data access permissions in ai agents.
\newblock \emph{arXiv preprint arXiv:2511.17959}, 2025.

\bibitem[Xiang et~al.(2026)Xiang, Zagieboylo, Ghosh, Kariyappa, Greshake, Xiao,
  Xiao, and Suh]{xiang2026architecting}
Chong Xiang, Drew Zagieboylo, Shaona Ghosh, Sanjay Kariyappa, Kai Greshake,
  Hanshen Xiao, Chaowei Xiao, and G~Edward Suh.
\newblock Architecting secure ai agents: Perspectives on system-level defenses
  against indirect prompt injection attacks.
\newblock \emph{arXiv preprint arXiv:2603.30016}, 2026.

\bibitem[Yan(2025)]{yan2025fault}
Boyang Yan.
\newblock Fault-tolerant sandboxing for ai coding agents: A transactional
  approach to safe autonomous execution.
\newblock \emph{arXiv preprint arXiv:2512.12806}, 2025.

\bibitem[Yang et~al.(2025)Yang, Shi, Ma, Liu, K{\"a}stner, and
  Wu]{yang2025prompts}
Chenyang Yang, Yike Shi, Qianou Ma, Michael~Xieyang Liu, Christian K{\"a}stner,
  and Tongshuang Wu.
\newblock What prompts don't say: Understanding and managing underspecification
  in llm prompts.
\newblock \emph{arXiv preprint arXiv:2505.13360}, 2025.

\bibitem[Yao et~al.(2022)Yao, Zhao, Yu, Du, Shafran, Narasimhan, and
  Cao]{yao2022react}
Shunyu Yao, Jeffrey Zhao, Dian Yu, Nan Du, Izhak Shafran, Karthik Narasimhan,
  and Yuan Cao.
\newblock React: Synergizing reasoning and acting in language models.
\newblock \emph{arXiv preprint arXiv:2210.03629}, 2022.

\bibitem[Zhan et~al.(2025)Zhan, Fang, Panchal, and Kang]{zhan2025adaptive}
Qiusi Zhan, Richard Fang, Henil~Shalin Panchal, and Daniel Kang.
\newblock Adaptive attacks break defenses against indirect prompt injection
  attacks on llm agents.
\newblock In \emph{Findings of the Association for Computational Linguistics:
  NAACL 2025}, pages 7101--7117, 2025.

\bibitem[Zheng et~al.(2023)Zheng, Chiang, Sheng, Zhuang, Wu, Zhuang, Lin, Li,
  Li, Xing, et~al.]{zheng2023judging}
Lianmin Zheng, Wei-Lin Chiang, Ying Sheng, Siyuan Zhuang, Zhanghao Wu, Yonghao
  Zhuang, Zi~Lin, Zhuohan Li, Dacheng Li, Eric Xing, et~al.
\newblock Judging llm-as-a-judge with mt-bench and chatbot arena.
\newblock \emph{Advances in neural information processing systems},
  36:\penalty0 46595--46623, 2023.

\bibitem[Zhong et~al.(2025)Zhong, Chen, Wang, McCall, Titzer, Miller, and
  Gibbons]{zhong2025rtbas}
Peter~Yong Zhong, Siyuan Chen, Ruiqi Wang, McKenna McCall, Ben~L Titzer,
  Heather Miller, and Phillip~B Gibbons.
\newblock Rtbas: Defending llm agents against prompt injection and privacy
  leakage.
\newblock \emph{arXiv preprint arXiv:2502.08966}, 2025.

\bibitem[Zhu et~al.(2025{\natexlab{a}})Zhu, Tseng, Vernik, Huang, Patil, Fang,
  and Popa]{zhu2025miniscope}
Jinhao Zhu, Kevin Tseng, Gil Vernik, Xiao Huang, Shishir~G Patil, Vivian Fang,
  and Raluca~Ada Popa.
\newblock Miniscope: A least privilege framework for authorizing tool calling
  agents.
\newblock \emph{arXiv preprint arXiv:2512.11147}, 2025{\natexlab{a}}.

\bibitem[Zhu et~al.(2025{\natexlab{b}})Zhu, Yang, Wang, Guo, and
  Wang]{zhu2025melon}
Kaijie Zhu, Xianjun Yang, Jindong Wang, Wenbo Guo, and William~Yang Wang.
\newblock Melon: Provable defense against indirect prompt injection attacks in
  ai agents.
\newblock \emph{arXiv preprint arXiv:2502.05174}, 2025{\natexlab{b}}.

\end{thebibliography}

\appendix
\section{SoK Methodology}
\label{app:methodology}

We define \ahi{} as any mechanism in which a human explicitly or implicitly
participates in a security-relevant decision made by, or about, an LLM agent.
This includes pre-deployment decisions (writing a policy, configuring a
sandbox), runtime decisions (approving a command, labeling a data source), and
implicit decisions (expressing intent through a task prompt). We exclude purely
automated defenses with no human in the loop and pure attack papers without a
defense contribution.

\noindent\textbf{Data sources.}
We drew from three sources. (i)~\emph{Academic papers}: we searched the
2022--\datayear proceedings of top-tier security (USENIX~Sec, S\&P, CCS, NDSS),
systems (OSDI, SOSP), ML (NeurIPS, ICML, ICLR), software-engineering (ICSE,
FSE), and NLP (ACL, EMNLP) venues, as well as arXiv, yielding \npapers papers.
(ii)~\emph{Production agent systems}: we catalogued \nsystems widely used coding,
computer-use, and enterprise agent platforms, drawing security details from
official documentation, engineering blogs, system cards, and public incident
reports. (iii)~\emph{Security plugins and tools}: we collected \nplugins
open-source and commercial tools marketed for agent security, including MCP
scanners, guardrail frameworks, and governance toolkits.

\noindent\textbf{Coding procedure.}
Two researchers independently performed open coding on a random 20\% sample of
papers and systems, then reconciled codes into an axial-coding schema. The
resulting five categories emerged as the minimal
covering set accounting for every observed \ahi{} mechanism. Each category was
refined with orthogonal analysis dimensions (e.g., granularity, enforcement
reliability, temporality). The remaining 80\% was coded by one researcher with
spot-checks from the second.

\section{Intent Alignment Category Analysis}
\label{app:category-analysis}

This appendix expands the per-category analysis summarized in the main text
(\S\ref{sec:sok}). For each \ahi{} category we walk through every analysis
dimension, report the distribution of papers, production systems, and plugins
across each dimension's sub-types, and surface cross-cutting patterns that
the main text could only allude to. Each subsection ends with one or two
takeaways and a short list of open problems that arise naturally from the
systematized landscape.

\subsection{C1: Policy Specification}
\label{app:c1}

Policy specification covers mechanisms by which a human writes rules that
constrain agent behavior. Across our corpus, 13 academic papers, 14 production
systems, and 8 plugins ship at least one policy-specification mechanism,
making C1 the most heavily represented category in academia and the second
most adopted in industry. We organize the design space along five orthogonal
dimensions: who authors the policy, the granularity of the rules, how the
rules are enforced at runtime, in what representation they are written, and
the temporality of the policy lifecycle. Table~\ref{tab:c1-dims} reports the
full coding.

\begin{table*}[t]
  \centering
  \small
  \caption{Policy Specification: every paper, production system, and plugin in our corpus that ships a policy-specification mechanism, characterized along five dimensions. Rows are color-coded: \colorbox{rowPaper}{\strut blue} for academic papers, \colorbox{rowSystem}{\strut green} for production systems, \colorbox{rowPlugin}{\strut orange} for plugins. For systems/plugins that ship multiple policy layers (e.g., \claudecode{} CLAUDE.md vs.\ Hooks vs.\ Blocklist), each layer is a separate row.}
  \label{tab:c1-dims}
  \setlength{\tabcolsep}{3pt}
  \begin{tabular}{@{}l c c c l c@{}}
    \toprule
    \textbf{Source} & \textbf{Author} & \textbf{Gran.} & \textbf{Enforc.} & \textbf{Representation} & \textbf{Temp.} \\
    \midrule
    \rowcolor{rowPaper} Progent~\cite{shi2025progent}               & \iconDeveloper            & \iconFine     & \iconDetEngine   & DSL                   & \iconStatic \\
    \rowcolor{rowPaper} Conseca~\cite{tsai2025contextual}              & \iconLLM+\iconEndUser     & \iconDynGran  & \iconLLMChecker  & NL$\to$formal         & \iconDynamic \\
    \rowcolor{rowPaper} AgentSpec~\cite{wang2026agentspec}          & \iconDeveloper            & \iconFine     & \iconDetEngine   & DSL                   & \iconStatic \\
    \rowcolor{rowPaper} CSAgent~\cite{gong2025csagent}              & \iconDeveloper            & \iconFine     & \iconDetEngine   & perm.\ model          & \iconStatic \\
    \rowcolor{rowPaper} AgentBound~\cite{buhler2025securing}      & \iconDeveloper (auto)     & \iconMedium   & \iconDetEngine   & perm.\ manif.         & \iconStatic \\
    \rowcolor{rowPaper} SEAgent~\cite{ji2026taming}                & \iconSecExpert            & \iconFine     & \iconDetEngine   & DSL (ABAC)            & \iconStatic \\
    \rowcolor{rowPaper} AGrail~\cite{luo2025agrail}                 & \iconAdmin                & \iconDynGran  & \iconLLMChecker  & NL$\to$formal         & \iconDynamic \\
    \rowcolor{rowPaper} DRIFT~\cite{li2025drift}                    & \iconLLM-gen              & \iconDynGran  & \iconDetEngine   & formal                & \iconDynamic \\
    \rowcolor{rowPaper} AgentGuardian~\cite{abaev2026agentguardian} & \iconLLM-learn            & \iconMedium   & \iconFwHook      & learned               & \iconDynamic \\
    \rowcolor{rowPaper} FATH~\cite{wang2024fath}                    & \iconDeveloper            & \iconMedium   & \iconFwHook      & \iconConfig           & \iconStatic \\
    \rowcolor{rowPaper} LLMZ+~\cite{pawelek2025llmzplus}            & \iconDeveloper            & \iconMedium   & \iconLLMChecker  & NL (guard)            & \iconStatic \\
    \rowcolor{rowPaper} MiniScope~\cite{zhu2025miniscope}           & \iconDeveloper            & \iconMedium   & \iconDetEngine   & perm.\ model          & \iconStatic \\
    \rowcolor{rowPaper} NeMo Guardrails~\cite{rebedea2023nemo}      & \iconDeveloper            & \iconFine     & \iconDetEngine   & DSL (Colang)          & \iconStatic \\
    \midrule
    \rowcolor{rowSystem} \claudecode{} CLAUDE.md~\cite{anthropic2025claudecode} & \iconEndUser & \iconCoarse--\iconMedium & \iconLLMSelf & NL              & \iconSession \\
    \rowcolor{rowSystem} \claudecode{} Hooks                          & \iconDeveloper           & \iconMedium   & \iconFwHook      & \iconCode (JS)        & \iconStatic \\
    \rowcolor{rowSystem} \claudecode{} Blocklist                      & \iconEndUser             & \iconMedium   & \iconDetEngine   & \iconConfig           & \iconStatic \\
    \rowcolor{rowSystem} \codex{} Rules~\cite{openai2025codex}        & \iconDeveloper           & \iconMedium   & \iconDetEngine   & \iconConfig           & \iconStatic \\
    \rowcolor{rowSystem} OpenClaw 3-layer~\cite{shan2026don}     & \iconDeveloper           & \iconMedium   & \iconDetEngine   & \iconConfig (JSON)    & \iconStatic \\
    \rowcolor{rowSystem} \cursorsys{} rules~\cite{cursor2025security} & \iconEndUser/\iconDeveloper & \iconMedium & \iconLLMSelf     & NL                    & \iconSession \\
    \rowcolor{rowSystem} \cursorsys{} admin policy                    & \iconAdmin               & \iconMedium   & \iconSysEnf      & \iconConfig           & \iconStatic \\
    \rowcolor{rowSystem} Windsurf \texttt{.codeiumignore}             & \iconDeveloper           & \iconMedium   & \iconSysEnf      & \iconConfig           & \iconStatic \\
    \rowcolor{rowSystem} Gemini CLI~\cite{google2025geminicli}        & \iconDeveloper           & \iconFine     & \iconDetEngine   & TOML                  & \iconStatic \\
    \rowcolor{rowSystem} Copilot CLI~\cite{github2025copilotcli}      & \iconEndUser             & \iconMedium   & \iconDetEngine   & \iconCmd flags        & \iconSession \\
    \rowcolor{rowSystem} LangChain~\cite{langchain2025}               & \iconDeveloper           & \iconFine     & \iconFwHook      & \iconCode (Py)        & \iconStatic \\
    \rowcolor{rowSystem} CrewAI~\cite{crewai2025}                     & \iconDeveloper           & \iconMedium   & \iconLLMFw       & \iconCode (Py)        & \iconStatic \\
    \rowcolor{rowSystem} Bedrock~\cite{aws2025bedrock}                & \iconAdmin               & \iconMedium   & \iconSysEnf      & \iconConfig           & \iconStatic \\
    \rowcolor{rowSystem} MS Agent FW~\cite{microsoft2025agentfw}      & \iconAdmin               & \iconFine     & \iconFwHook      & \iconConfig           & \iconStatic \\
    \midrule
    \rowcolor{rowPlugin} Intercept~\cite{intercept2025}               & \iconDeveloper           & \iconMedium--\iconFine & \iconDetEngine & YAML            & \iconStatic \\
    \rowcolor{rowPlugin} assay                                       & \iconDeveloper           & \iconMedium   & \iconDetEngine   & policy                & \iconStatic \\
    \rowcolor{rowPlugin} GUARDRAIL                                   & \iconDeveloper           & \iconMedium   & \iconDetEngine   & \iconConfig           & \iconStatic \\
    \rowcolor{rowPlugin} Guardrails AI~\cite{guardrailsai2025}        & \iconDeveloper           & \iconFine     & \iconFwHook      & \iconCode (Py)        & \iconStatic \\
    \rowcolor{rowPlugin} NeMo Guardrails (OSS)~\cite{rebedea2023nemo} & \iconDeveloper           & \iconFine     & \iconDetEngine   & DSL (Colang)          & \iconStatic \\
    \rowcolor{rowPlugin} OpenAI Guardrails                           & \iconDeveloper           & \iconMedium   & \iconFwHook      & \iconCode (Py)        & \iconStatic \\
    \rowcolor{rowPlugin} cursor-security-rules                       & \iconEndUser             & \iconCoarse   & \iconLLMSelf     & NL                    & \iconSession \\
    \rowcolor{rowPlugin} SecureClaw~\cite{secureclaw2025}             & \iconDeveloper           & \iconMedium   & \iconFwHook      & \iconConfig (OWASP)   & \iconStatic \\
    \bottomrule
  \end{tabular}

  \vspace{4pt}
  \raggedright\small
  \textbf{Legend ---}
  Author: \iconDeveloper\,developer, \iconEndUser\,end user, \iconAdmin\,admin, \iconSecExpert\,sec.\ expert, \iconLLM\,LLM.\quad
  Gran.: \iconCoarse\,coarse, \iconMedium\,med., \iconFine\,fine, \iconDynGran\,dynamic.\quad
  Enforc.: \iconDetEngine\,det.\ engine, \iconLLMChecker\,LLM checker, \iconLLMSelf\,LLM self-compl., \iconFwHook\,fw.\ hook, \iconSysEnf\,sys.\ enforc., \iconLLMFw\,LLM+fw.\quad
  Repr.\ entities: \iconCode\,code, \iconConfig\,config, \iconCmd\,CLI.\quad
  Temp.: \iconStatic\,static, \iconDynamic\,dynamic, \iconSession\,session.
\end{table*}

\noindent\textbf{Author.}
The author dimension records who is expected to write the policy, and its
distribution diverges sharply between academia and production. End users
authoring free-form natural-language constraints account for 4 production
policy layers (\claudecode{}'s
\texttt{CLAUDE.md}~\cite{anthropic2025claudecode}, \claudecode{}'s blocklist,
\cursorsys{}'s rules files~\cite{cursor2025security}, Copilot CLI session-wide
flags~\cite{github2025copilotcli}) and 1 plugin (\texttt{cursor-security-rules}),
but appear in 0 of 13 papers; every academic proposal assumes a more privileged
author. Developers and admins using a configuration language, a permission
manifest, or general-purpose code dominate the corpus, accounting for 9 of 13
papers (Progent~\cite{shi2025progent}, AgentSpec~\cite{wang2026agentspec},
CSAgent~\cite{gong2025csagent}, AgentBound~\cite{buhler2025securing},
FATH~\cite{wang2024fath}, LLMZ+~\cite{pawelek2025llmzplus},
MiniScope~\cite{zhu2025miniscope}, NeMo Guardrails~\cite{rebedea2023nemo}) and
the majority of production layers (\codex{}~\cite{openai2025codex},
OpenClaw~\cite{shan2026don}, Gemini CLI~\cite{google2025geminicli},
LangChain~\cite{langchain2025}, CrewAI~\cite{crewai2025}, Windsurf,
plus 6 plugins). Security experts authoring MAC or ABAC rules
(SEAgent~\cite{ji2026taming}, the advanced configurations of NeMo Guardrails)
appear in only 1 paper plus the \cursorsys{} admin and Bedrock admin
consoles~\cite{aws2025bedrock}. Finally, LLM-generated policy with
human review (Conseca~\cite{tsai2025contextual}, AGrail~\cite{luo2025agrail},
DRIFT~\cite{li2025drift}, AgentGuardian~\cite{abaev2026agentguardian})
accounts for 4 papers and 0 production layers.

\noindent\textbf{Granularity.}
Granularity measures how narrow a policy rule's scope is. Coarse global rules
applying to all actions of a class are rare academically (0 of 13 papers) but
present in 3 of 14 systems via sandbox-mode flags
(\codex{} \texttt{workspace-write} versus \texttt{danger-full-access},
\claudecode{} sandbox on/off) and the \texttt{cursor-security-rules}
plugin. Medium per-tool or per-resource rules are the modal sub-type, used by
6 papers and 9 of 14 production policy layers
(\claudecode{}'s blocklist matching \texttt{curl}/\texttt{wget}/etc.,
OpenClaw \texttt{agents.list[].tools.allow/deny}, Copilot CLI
\texttt{--allow-tool} flags, AgentBound MCP-server permissions, FATH, LLMZ+,
MiniScope, AgentGuardian). Fine per-action or per-parameter rules with
fallback behavior dominate academic work (7 papers including Progent's DSL,
AgentSpec's trigger plus predicate plus enforcement triples, CSAgent,
SEAgent, NeMo Guardrails Colang, Gemini CLI TOML, LangChain middleware) but
appear in only 3 of 14 production systems and 4 plugins. Dynamic
context-dependent rules generated at runtime are the academic frontier
(Conseca, DRIFT, AGrail) with no production deployment.

\noindent\textbf{Enforcement mechanism.}
Enforcement determines how reliably a written policy actually constrains the
agent, and is the single most important axis along which academia and
industry diverge. We classify mechanisms along a reliability spectrum.
LLM self-compliance, where the policy is embedded in a system prompt and the
backbone LLM is expected to honor it, anchors 5 of 14 production policy
layers (\claudecode{}'s \texttt{CLAUDE.md}, \cursorsys{} rules files, embedded
skill instructions, \devin{}'s planning instructions~\cite{devin2025cognition})
but no academic proposal: a prompt injection that contradicts the policy can
directly subvert this layer. LLM-as-checker designs, where a second model
audits outputs, appear in 4 papers (Conseca, AGrail, LLMZ+, AgentGuardian) and
1 production system (CrewAI's LLM-as-judge guardrails); the checker improves
on self-compliance but remains a soft target. Deterministic engines, where an
independent runtime interpreter applies the policy without calling the LLM,
dominate academia (7 of 13 papers: Progent, AgentSpec, CSAgent, AgentBound,
SEAgent, DRIFT, MiniScope, NeMo Guardrails) and reach production via
\claudecode{}'s blocklist, \codex{} rules, OpenClaw, the Gemini CLI TOML
engine, Copilot CLI, plus the four deterministic plugins
(Intercept~\cite{intercept2025}, \texttt{assay}, GUARDRAIL, NeMo Guardrails OSS).
Framework hooks (Guardrails AI~\cite{guardrailsai2025} validators, OpenAI
guardrails, LangChain \texttt{HumanInTheLoopMiddleware}, \claudecode{} Hooks,
MS Agent Framework proxy~\cite{microsoft2025agentfw}, SecureClaw~\cite{secureclaw2025},
plus FATH and AgentGuardian on the academic side) sit one rung above engines
in deployment ease and one rung below in dispatch determinism.
OS or system enforcement (macOS Seatbelt and Linux
Bubblewrap in \claudecode{}~\cite{anthropic2025claudecodesandbox},
\cursorsys{} admin, Windsurf \texttt{.codeiumignore},
Bedrock IAM) is the highest-reliability sub-type but rarely appears as the
primary mechanism in academic policy work, which treats OS isolation as
orthogonal to the policy engine itself.

\noindent\textbf{Representation.}
Representation captures the syntactic form of the policy and tracks the
author dimension almost one-to-one. Natural language, as in \texttt{CLAUDE.md},
user prompt addenda, embedded skills, and \devin{} planning instructions, is
zero-friction to write but inherits all the reliability problems of LLM
self-compliance. Declarative configuration files with a fixed schema
(Intercept YAML, Gemini CLI TOML, OpenClaw JSON, \texttt{.codeiumignore},
\codex{} rules, Bedrock console settings) are the modal production form.
Domain-specific languages designed for security (Progent's DSL, AgentSpec's
trigger plus predicate plus enforcement language, NeMo Guardrails Colang,
SEAgent's ABAC language) compile to deterministic engines and are the modal
academic form. Code-level policy in general-purpose languages (Guardrails AI
Python validators, OpenAI guardrail functions, \claudecode{} Hooks in
JavaScript, LangChain middleware in Python) is strictly more expressive than
configuration but requires programmer authorship. Permission models or
manifests (AgentBound's Android-style manifest, MiniScope's hierarchical
permissions, Bedrock IAM condition keys) trade expressiveness for
auditability.

\noindent\textbf{Temporality.}
Temporality records when the policy is fixed relative to agent execution.
Static policies defined before deployment dominate the corpus: 8 of 13 papers
and 11 of 14 production policy layers are static. Session-level policies
specified at session start (\texttt{CLAUDE.md}, \cursorsys{} rules, Copilot CLI
session-wide permissions) are the natural compromise for end-user authoring,
cheap to update but inheriting the same self-compliance reliability concerns.
Dynamic policies generated or rewritten during execution (Conseca's
just-in-time contextual policies, DRIFT's query-derived rules, AGrail's
lifelong learning, AgentGuardian's behavior-derived policies) account for 4
papers and 0 production systems.

\noindent\textbf{Cross-cutting observations.}
The five dimensions are highly correlated rather than independent. Author and
representation move together: end users overwhelmingly choose natural
language, developers choose configuration or code, security experts choose
DSLs, LLMs emit formal forms. Author also predicts enforcement: end-user
policy collapses to LLM self-compliance, developer policy reaches
deterministic engines. The most reliable enforcement layers (deterministic
engine, OS) are written in the least accessible representations (DSL, IAM
manifests) and authored by the most privileged actors. The asymmetry
produces a structural usability cliff: an end user cannot reach the
high-reliability layers without first becoming a developer. A second pattern
is the academia-industry mismatch on temporality: of the 4 papers that
move beyond static policy, all rely on LLM-as-checker or learned policies,
while production systems with dynamic temporality essentially do not exist;
the static policies that dominate production remain reliable in the
deterministic-engine sense but age poorly as the agent's tool surface grows.

\takeaways{\label{tk:app-c1-correlation}%
The five C1 dimensions are not independent: author, representation, and
enforcement reliability co-vary. End-user policy collapses to natural language
and LLM self-compliance, developer policy reaches deterministic engines, and
the high-reliability tail (DSL plus OS enforcement) is unreachable without
programmer or admin privileges. The structural consequence is that end users
inherit the lowest-reliability enforcement by default, with no on-ramp.
}

\subsection{C2: Trust and Data Labeling}
\label{app:c2}

Trust and data labeling covers mechanisms by which a human (or a
human-configured proxy) annotates data, tools, agents, or communication
channels with a trust or integrity label, and the system propagates that label
through agent execution to gate downstream decisions. Across our corpus, 8
academic papers ship a trust-labeling design but \emph{zero} production
systems and \emph{zero} plugins do, making C2 the largest academia-industry
gap we observed. We organize the design space along six orthogonal dimensions:
what is labeled, the type of label, propagation rules, the labeling subject,
how labels are enforced, and the completeness requirement.
Table~\ref{tab:c2-dims} reports the full coding.

\begin{table*}[t]
  \centering
  \small
  \caption{Trust \& Data Labeling: every paper in our corpus that ships a trust/data labeling mechanism, characterized along six dimensions. No production system or plugin currently exposes trust labeling as a user-facing mechanism, so the table contains academic sources only (rows in \colorbox{rowPaper}{\strut blue}).}
  \label{tab:c2-dims}
  \setlength{\tabcolsep}{3pt}
  \begin{tabular}{@{}l l c c c c l@{}}
    \toprule
    \textbf{Source} & \textbf{Label obj.} & \textbf{Label type} & \textbf{Propag.} & \textbf{Subject} & \textbf{Enforc.} & \textbf{Complete.} \\
    \midrule
    \rowcolor{rowPaper} Fides~\cite{costa2025securing}                    & \iconData src               & \iconLattice   & \iconDetTaint  & \iconSecExpert  & \iconIFC          & full \\
    \rowcolor{rowPaper} SafeFlow~\cite{li2025safeflow}                 & \iconData                   & \iconLattice   & \iconDetTaint  & \iconSecExpert  & \iconIFC          & full \\
    \rowcolor{rowPaper} CaMeL~\cite{debenedetti2025defeating}              & \iconAgent/\iconTool        & \iconCapBased  & \iconLLMReason & \iconEndUser    & \iconCapCheck     & partial (deny) \\
    \rowcolor{rowPaper} PFI~\cite{kim2025prompt}                          & \iconAgent                  & \iconBinaryTrust & \iconManual  & \iconEndUser    & \iconIsolated     & partial (untrust) \\
    \rowcolor{rowPaper} PermissiveIFC~\cite{siddiqui2024permissive} & \iconData flow              & \iconLattice   & \iconSelective & \iconDeveloper  & \iconFilter       & incremental \\
    \rowcolor{rowPaper} f-secure-IFC~\cite{wu2024system}           & input ch.                   & \iconBinaryTrust & \iconDetTaint & \iconDeveloper & \iconInputFilter  & partial (untrust) \\
    \rowcolor{rowPaper} AgentArmor~\cite{wang2025agentarmor}           & \iconTool                   & \iconCapBased  & \iconManual    & \iconDeveloper  & \iconCapCheck     & full (registry) \\
    \rowcolor{rowPaper} VerifiablySafe~\cite{doshi2026towards}  & \iconTool (\iconMCP)        & \iconMultiLevel & \iconManual   & \iconDeveloper  & \iconFormalVerif  & full \\
    \bottomrule
  \end{tabular}

  \vspace{4pt}
  \raggedright\small
  \textbf{Legend ---}
  Label type: \iconLattice\,lattice, \iconCapBased\,capability, \iconBinaryTrust\,binary, \iconMultiLevel\,multi-level.\quad
  Propag.: \iconDetTaint\,det.\ taint, \iconLLMReason\,LLM reason., \iconManual\,manual, \iconSelective\,selective.\quad
  Subject: \iconSecExpert\,sec.\ expert, \iconEndUser\,end user, \iconDeveloper\,developer.\quad
  Enforc.: \iconIFC\,IFC, \iconCapCheck\,cap.\ check, \iconIsolated\,isolation, \iconFilter\,filter, \iconInputFilter\,input filter, \iconFormalVerif\,formal verif.\quad
  Entities: \iconData\,data, \iconAgent\,agent, \iconTool\,tool, \iconMCP\,MCP.
\end{table*}

\noindent\textbf{Label object.}
Academic designs split across four label-object sub-types. Data and input
sources are labeled in 4 of 8 papers: Fides~\cite{costa2025securing} attaches
dual confidentiality plus integrity labels to every input;
SafeFlow~\cite{li2025safeflow} attaches lattice-ordered labels to data;
f-secure-IFC~\cite{wu2024system} labels input channels as trusted or
untrusted; PermissiveIFC~\cite{siddiqui2024permissive} labels data flow
paths. Tool or API metadata is labeled in 2 papers
(AgentArmor~\cite{wang2025agentarmor} property registry,
VerifiablySafe~\cite{doshi2026towards} structured
$\langle$capability, confidentiality, trust$\rangle$ labels for MCP tools).
Agents or actors are labeled in 2 papers (PFI~\cite{kim2025prompt} divides agents
into trusted and untrusted partitions; CaMeL~\cite{debenedetti2025defeating}
associates capabilities with each agent or tool). Channel-level labels appear
in PermissiveIFC and f-secure-IFC, which ride labels on the path rather than
the endpoints.

\noindent\textbf{Label type.}
The algebraic structure of the label space ranges from minimal to expressive.
Binary trusted/untrusted labels (PFI, f-secure-IFC) discard ordering
information needed for nuanced flow rules but are easy to reason about.
Multi-level structured tuples (VerifiablySafe's
$\langle$capability, confidentiality, trust$\rangle$) compose multiple
attribute axes into a single label. Lattice-based labels (Fides, SafeFlow)
join confidentiality and integrity into a partial order with meet and join,
supporting principled propagation in the classical Denning sense.
Capability-based labels (CaMeL, AgentArmor) record what an entity can do
rather than where it sits in a hierarchy.

\noindent\textbf{Propagation.}
Propagation rules determine how labels flow through agent execution. Manual
exhaustive labeling, where every entity is labeled by hand and labels do not
flow (PFI), keeps reasoning local at the cost of high labeling effort.
Deterministic taint, where labels follow data through every operation in the
classical sense (Fides, SafeFlow), is sound but over-tainting is common.
Selective propagation, the central contribution of PermissiveIFC, propagates
a label only when the underlying datum actually influences the output and
matches the deterministic result in 85\%+ of cases while shrinking the
over-tainted set considerably; this is the only sub-type that explicitly
trades soundness for utility. LLM-mediated propagation (CaMeL) lets the model
honor capability constraints when chaining tools, lowering the user-facing
burden but weakening soundness relative to deterministic taint.

\noindent\textbf{Subject.}
The subject dimension records who assigns labels. End-user labeling
(CaMeL, PFI) assumes the user can articulate which tools or agents are
trusted, a strong assumption. Developer or admin labeling at deployment time
(AgentArmor's tool property registry, VerifiablySafe's MCP tool labels,
f-secure-IFC's monitor configuration) is the modal academic choice.
Security-expert labeling assuming IFC fluency (Fides, SafeFlow, the advanced
VerifiablySafe configurations) raises the expertise bar but enables
formally-grounded designs. Semi-automatic labeling, where the system
synthesizes and the human verifies (PermissiveIFC's selective rules,
AgentBound's 80.9\% inference from source code), is the lowest-burden
sub-type, although AgentBound itself sits primarily in C1.

\noindent\textbf{Enforcement.}
How labels actually drive a security decision. IFC engines blocking flows
(Fides, SafeFlow) prevent low-integrity data from reaching high-integrity
sinks; Fides reports blocking every prompt-injection benchmark attack in its
evaluation. Capability checks before each call (CaMeL) achieve provable
security on 77\% of evaluated tasks. Isolated execution
(PFI runs untrusted agents in environments that cannot reach the trusted
agent's plugins; f-secure-IFC filters untrusted input out of planning)
separates label classes physically or logically. Filter or downgrade
(PermissiveIFC) treats untrusted data as a softer signal rather than blocking
outright. Formal verification (VerifiablySafe with Alloy) checks the design
itself against an explicit security property.

\noindent\textbf{Completeness.}
Completeness records how exhaustive the labeling has to be. Full labeling,
where every entity must be labeled before the system runs (Fides, SafeFlow,
VerifiablySafe), imposes a high up-front burden. Partial plus default-deny,
where unlabeled entities receive a default (CaMeL, PFI default to deny or
untrusted), reduces burden but shifts the failure mode to silent
over-blocking. Incremental labeling, where only inputs and sinks are labeled
and the rest is inferred (PermissiveIFC, AgentBound), is the
lowest-burden academic design point and the natural place adoption would
start if cost were the only barrier.

\noindent\textbf{Cross-cutting observations.}
The most striking pattern in C2 is the absence of any production adoption.
The cost-based hypothesis predicts that low-burden variants should appear
first, yet PermissiveIFC's incremental labeling and CaMeL's default-deny have
zero industry uptake despite predating the higher-burden variants. The
closest production analogues are \texttt{mcp-scan}~\cite{mcpscan2025} and
\texttt{mcpserver-audit}, which generate one-time risk reports users consult
before installation: this is C2 reframed as a C3 review decision rather than
a propagating label. Salesforce's Einstein Trust
Layer~\cite{salesforce2025agentforce} ships internal data labels at the
platform layer, but the labels are invisible to the user and do not satisfy
our \ahi{} criterion. A second observation is that label-object choice
predicts enforcement strategy: data-source labels lead to IFC engines,
tool-level labels to capability checks, agent-level labels to isolation,
flow-level labels to filtering. The diversity is welcome from a research
perspective but compounds the adoption barrier, because committing to a
mental model implicitly commits to one enforcement implementation.

\takeaways{\label{tk:app-c2-mentalmodel}%
8 academic papers, 0 production systems: trust labeling is the largest
academia-industry gap in our corpus. The cost-based explanation does not fit
because the lowest-burden academic variants (PermissiveIFC, CaMeL with
default-deny) also have zero adoption. The remaining hypothesis is that
end users do not natively reason about data sources or tool capabilities as
carriers of trust, even when they happily reason about whether an
\emph{individual action} is risky.
}

\subsection{C3: Runtime Approval}
\label{app:c3}

Runtime approval is the only category that pauses agent execution to defer a
decision to the human. Across our corpus, 5 academic papers, 15 production
systems, and 4 plugins ship at least one runtime approval mechanism, making
C3 the most widely deployed category in industry alongside scope
configuration. We organize the design
space along six dimensions: what is approved, the granularity of the
approval, the trigger condition, the response options offered to the user,
fatigue-mitigation strategies, and how much information about the action is
presented at decision time. Table~\ref{tab:c3-dims} reports the full coding.

\begin{table*}[t]
  \centering
  \small
  \caption{Runtime Approval: every paper, production system, and plugin in our corpus that ships a runtime-approval mechanism, characterized along six dimensions. Rows are color-coded: \colorbox{rowPaper}{\strut blue} for academic papers, \colorbox{rowSystem}{\strut green} for production systems, \colorbox{rowPlugin}{\strut orange} for plugins.}
  \label{tab:c3-dims}
  \setlength{\tabcolsep}{3pt}
  \begin{tabular}{@{}l l l l c l l@{}}
    \toprule
    \textbf{Source} & \textbf{Object} & \textbf{Gran.} & \textbf{Trigger} & \textbf{Opts.} & \textbf{Fatigue mitig.} & \textbf{Info shown} \\
    \midrule
    \rowcolor{rowPaper} IsolateGPT~\cite{wu2024isolategpt}        & cross-app          & /\iconTool        & cross-app           & \iconBinary    & ---                  & raw \\
    \rowcolor{rowPaper} ACE~\cite{li2025ace}                       & \iconPlan          & /\iconPlan        & auto                & \iconBinary    & plan-level           & \iconPlan \\
    \rowcolor{rowPaper} RTBAS~\cite{zhong2025rtbas}                & \iconCmd           & risk-adapt.  & high-risk           & \iconBinary    & risk screen          & \iconCmd+risk \\
    \rowcolor{rowPaper} OpenClaw HITL~\cite{shan2026don}      & risky op           & /\iconCmd         & high-risk           & \iconBinary    & ---                  & \iconCmd \\
    \rowcolor{rowPaper} AutoPermissions~\cite{wu2025towards} & \iconData        & /\iconData         & \iconData access    & \iconBinaryMem & ML pred.\ (85\%)     & \iconCmd+ctx \\
    \midrule
    \rowcolor{rowSystem} \claudecode~\cite{anthropic2025claudecode}    & \iconCmd+\iconFile & /\iconCmd       & \iconEndUser tier   & \iconBinaryMem & always+\iconEnv      & \iconCmd/\iconDiff \\
    \rowcolor{rowSystem} \codex~\cite{openai2025codex}                  & \iconCmd          & risk-adapt.  & \iconEnv            & \iconBinary    & \iconEnv auto         & \iconCmd \\
    \rowcolor{rowSystem} Manus                                         & \iconCmd          & /\iconCmd         & all                 & \iconBinaryMem & always-allow          & minimal \\
    \rowcolor{rowSystem} OpenClaw                                      & \iconTool         & /\iconTool        & \iconConfig         & \iconBinary    & ---                   & \iconTool \\
    \rowcolor{rowSystem} \cursorsys~\cite{cursor2025security}           & \iconCmd+\iconMCP & /\iconCmd         & default all         & \iconBinary    & auto-run              & \iconCmd \\
    \rowcolor{rowSystem} Windsurf                                      & \iconCmd          & /\iconCmd         & \iconConfig         & \iconBinary    & disable auto          & \iconCmd \\
    \rowcolor{rowSystem} \devin~\cite{devin2025cognition}               & \iconPlan+\iconPR & /\iconPlan+\iconPR     & 2 checkpts          & \iconBinaryFb  & plan-level            & \iconPlan+\iconDiff \\
    \rowcolor{rowSystem} GitHub Copilot                                & \iconCmd          & risk-adapt.  & \iconEnv            & \iconBinary    & \iconEnv auto         & \iconCmd \\
    \rowcolor{rowSystem} Gemini CLI~\cite{google2025geminicli}          & \iconCmd+priv.    & /\iconCmd         & default all         & \iconBinary    & policy auto           & \iconCmd+perm \\
    \rowcolor{rowSystem} Copilot CLI~\cite{github2025copilotcli}        & \iconCmd+\iconDir & /\iconCmd         & new \iconDir/\iconCmd & \iconBinaryMem & session+YOLO        & \iconCmd \\
    \rowcolor{rowSystem} LangChain~\cite{langchain2025}                 & \iconTool         & /\iconTool        & interrupt           & \iconThreeWay  & config.\ auto         & \iconTool+params \\
    \rowcolor{rowSystem} ChatGPT                                       & \iconAPI          & /action      & ext.\ \iconAPI      & \iconBinary    & ---                   & minimal \\
    \rowcolor{rowSystem} MS Agent FW~\cite{microsoft2025agentfw}        & wkfl.\ step       & /checkpoint  & low conf.           & \iconBinaryFb  & conf.\ escalation     & multi-layer \\
    \midrule
    \rowcolor{rowPlugin} mcp-guardian~\cite{mcpguardian2025}            & \iconMCP          & /\iconTool        & \iconConfig         & \iconBinary    & policy pass           & \iconMCP \\
    \rowcolor{rowPlugin} safety-net~\cite{safetynet2025}                & danger.\ \iconCmd & /\iconCmd         & danger.\ match      & \iconBinary    & danger.\ only         & \iconCmd+reason \\
    \rowcolor{rowPlugin} Claude Code Hooks                             & danger.\ \iconCmd  & /\iconCmd         & danger.\ match      & \iconBinary    & danger.\ only         & \iconCmd \\
    \rowcolor{rowPlugin} Trail of Bits Skills~\cite{trailofbits2025skills} & \iconReview     & /task        & \iconEndUser invoc. & \iconAdvisory  & on-demand             & \iconReview \\
    \bottomrule
  \end{tabular}

  \vspace{4pt}
  \raggedright\small
  \textbf{Legend ---}
  Opts.: \iconBinary\,binary, \iconBinaryMem\,binary+memory, \iconBinaryFb\,binary+feedback, \iconThreeWay\,three-way (edit), \iconAdvisory\,advisory.\quad
  Entities: \iconCmd\,command/CLI, \iconTool\,tool, \iconMCP\,MCP, \iconAPI\,API, \iconFile\,file, \iconDir\,directory, \iconEnv\,sandbox/exec.\ env, \iconPlan\,plan, \iconPR\,pull request, \iconDiff\,diff, \iconData\,data, \iconReview\,code review, \iconConfig\,config, \iconEndUser\,end user.\quad
  Granularity \emph{/X} reads ``per-X''.
\end{table*}

\noindent\textbf{Approval object.}
What the user is asked to allow or reject. Single commands or calls dominate
production: most of the 15 systems prompt on every Bash command (\claudecode{}),
every terminal command (\manus{}), or every external API call (ChatGPT
Actions). Batched operations bundling multiple related actions for a single
decision appear in IsolateGPT~\cite{wu2024isolategpt}, motivated by cutting
prompt count without losing per-action visibility. Plan-level approval
formalized by ACE~\cite{li2025ace} replaces step-by-step prompts with a
single decision over an abstract plan; \devin{} ships a similar plan
checkpoint followed by a separate PR checkpoint. File or code changes
reviewed as a diff appear in \claudecode{}'s file edit prompts and
\devin{}'s PR review. Permission elevation prompts (Gemini CLI's
expansion dialog, \codex{}'s sandbox-boundary expansion) ask the user to
grant a specific scope expansion in response to an agent request.

\noindent\textbf{Granularity.}
The resolution of the approval boundary. Per-command granularity is the
default for \claudecode{}, \manus{}, Copilot CLI, \cursorsys{}, Windsurf, and
the four C3 plugins (\texttt{mcp-guardian}~\cite{mcpguardian2025},
\texttt{safety-net}~\cite{safetynet2025}, \texttt{hooks-mastery},
\texttt{claude-code-hooks}). Per-tool-call granularity, keyed to the abstract
tool rather than the literal command, is used by LangChain's
\texttt{interrupt()}, ChatGPT Actions, and the default mode of
\texttt{mcp-guardian}. Per-plan or checkpoint granularity is the academic
choice (ACE, \devin{}). Per-session granularity collapses many decisions into
a single up-front grant (Copilot CLI session-wide permissions,
\claudecode{}'s ``always allow''). Risk-adaptive granularity adjusts to
estimated risk: RTBAS~\cite{zhong2025rtbas} reports 2\% utility loss while
preventing all targeted attacks, the strongest published result; \codex{}
approximates this pattern by auto-approving sandboxed operations and only
prompting on sandbox-boundary crossings.

\noindent\textbf{Trigger condition.}
When the system actually pauses. The default-ask trigger that prompts on
every action appears in Gemini CLI's restricted mode and \claudecode{}'s
default mode. Sandbox-boundary triggers, where operations inside a sandbox
auto-approve and only those crossing the boundary interrupt, appear in
\codex{}, GitHub Copilot agent mode, and sandboxed \claudecode{}. High-risk
triggers, where the system estimates risk and only fires above a threshold,
are used by RTBAS and \cursorsys{}. Pattern-match triggers fire on
predefined dangerous patterns (\claudecode{}'s blocklist, the four hook-style
plugins). User-defined rules let the user configure which actions ask, allow,
or deny (\claudecode{}'s permission tiers, \texttt{mcp-guardian}'s policy file,
Copilot CLI \texttt{--allow-tool}/\texttt{--deny-tool}). Confidence-driven
escalation, where the agent itself escalates when its confidence falls below
a threshold, is supported as a first-class proxy policy by Microsoft Agent
Framework.

\noindent\textbf{Approval options.}
What the user can choose at the approval point. Binary allow/deny is the
minimal interface (ChatGPT Actions, \manus{} default). Binary plus memory,
where the user can pin the choice for the rest of the session or permanently,
is the most common production sub-type (\claudecode{}'s ``always allow'',
Copilot CLI session-wide grants, \manus{}'s ``Always Allow''); each click
expands the trusted set monotonically. Three-way (approve, edit, reject) lets
the user modify the action's parameters before approving and is exposed by
LangChain's interrupt API. Tiered options (\cursorsys{} admin: allow, warn,
require step-up, deny) map different approval levels to different post-approval
permissions. Approve-plus-feedback (\devin{} plan reviews, Microsoft Agent
Framework checkpoint reviews) lets the user attach comments or
modifications.

\noindent\textbf{Fatigue mitigation.}
Strategies for reducing how often the user is interrupted. Always-allow
memory permanently authorizes future instances of the same action and is the
most common production strategy and the one with the worst long-term
security profile, since each click monotonically expands the trusted surface.
Sandbox auto-approval delegates the decision to a sandbox boundary; Anthropic
reports that sandbox-based auto-approval cuts \claudecode{} prompt count by
84\%, and \codex{} and GitHub Copilot follow the same pattern. Risk-adaptive
screening (RTBAS) is the only published design with both strong utility and
strong attack-prevention results, but no production system we surveyed has
adopted it. Plan-level approval (ACE, \devin{}) replaces approving each step
with approving the plan. Session-wide authorization (Copilot CLI session
permissions, \cursorsys{} Auto-Run mode) collapses the session into a single
up-front grant. YOLO or full auto-run (Copilot CLI YOLO mode, \cursorsys{}
Auto-Run) skips all approvals: the lowest-friction and lowest-security point
in the design space.

\noindent\textbf{Information presentation.}
Decision quality ultimately depends on what the user sees. Minimal-information
prompts (``Allow this action?'' with no context, as in ChatGPT Actions and
\manus{} default) leave the user essentially guessing. Raw command or call
text (\claudecode{} default, \texttt{mcp-guardian}, most hook-style plugins)
allows technical users to judge but typically not non-technical ones.
Command plus risk explanation (RTBAS, Gemini CLI permission expansion,
Microsoft Agent Framework checkpoints) explains why the action is considered
risky. Diff previews (\claudecode{} file edit, \devin{} PR review) show file
changes in full detail. Abstract plans (ACE, \devin{}) trade fine-grained
visibility for cognitive accessibility. Multi-layer expandable presentations
(\cursorsys{} admin, GitHub Copilot agent mode) default to a summary with
detail on demand and are the highest-resolution presentation we observed.

\noindent\textbf{Cross-cutting observations.}
The runtime-approval design space exhibits a single cleanly articulated
tradeoff between fatigue and security, mediated by the fatigue-mitigation
choice. Always-allow strategies optimize for fatigue but silently widen the
trusted surface; sandbox auto-approval keeps the surface bounded but only
works when the sandbox is correctly configured (a C4 problem); risk-adaptive
screening is the only mechanism that claims both low fatigue and high
security, and it is the only one with no production adoption. A second
pattern is the loose coupling between approval object and information
presentation: diff previews and abstract plans are the information-richest
presentations, but they are paired with file edits and plan-level approvals
respectively, not with the per-command approvals that dominate production;
the mass of production approval prompts (raw command on a single line) is
the case where information is shallowest.

\takeaways{\label{tk:app-c3-tradeoff}%
Every production fatigue-mitigation strategy either expands the trusted
surface (always-allow, session-wide) or relies on a correctly configured
orthogonal mechanism (sandbox). Risk-adaptive screening occupies the
fourth quadrant of high security plus low fatigue but remains academic.
The most information-rich presentations (diff, abstract plan) are paired
with the rarest approval objects, not the per-command default that
dominates daily use.
}

\subsection{C4: Scope and Boundary Configuration}
\label{app:c4}

Scope and boundary configuration covers mechanisms by which a human defines
which tools, files, networks, and execution environments the agent can
touch. Across our corpus, 4 academic papers, 16 production systems, and 6
plugins ship a scope-configuration mechanism, making C4 (with C3) the most
common form of \ahi{} when counted at the category level. We organize the
design space along six dimensions: what is bounded, how the boundary is
defined, the granularity of the boundary, the enforcement layer, the default
stance, and how the boundary changes over the session.
Table~\ref{tab:c4-dims} reports the full coding.

\begin{table*}[t]
  \centering
  \small
  \caption{Scope \& Boundary Configuration: every paper, production system, and plugin in our corpus that ships a scope/boundary mechanism, characterized along six dimensions. Rows are color-coded: \colorbox{rowPaper}{\strut blue} for academic papers, \colorbox{rowSystem}{\strut green} for production systems, \colorbox{rowPlugin}{\strut orange} for plugins.}
  \label{tab:c4-dims}
  \setlength{\tabcolsep}{3pt}
  \begin{tabular}{@{}l l l l c c c@{}}
    \toprule
    \textbf{Source} & \textbf{Object} & \textbf{Definition} & \textbf{Gran.} & \textbf{Enforc.} & \textbf{Default} & \textbf{Mutab.} \\
    \midrule
    \rowcolor{rowPaper} IsolateGPT~\cite{wu2024isolategpt}              & \iconEnv/app                & arch.\ (hub)        & /app           & \iconFramework               & \iconClosed  & \iconImmutable \\
    \rowcolor{rowPaper} MiniScope~\cite{zhu2025miniscope}               & \iconTool/\iconAPI          & perm.\ hier.        & /tool          & \iconFramework               & \iconClosed  & \iconImmutable \\
    \rowcolor{rowPaper} FaultTolerantSandbox~\cite{yan2025fault} & \iconEnv+\iconFile   & policy \iconConfig  & /resource      & \iconOS                      & \iconClosed  & \iconImmutable \\
    \rowcolor{rowPaper} AgentBound~\cite{buhler2025securing}          & \iconTool (\iconMCP)        & perm.\ manif.       & /tool          & \iconFramework               & \iconClosed  & \iconImmutable \\
    \midrule
    \rowcolor{rowSystem} \claudecode~\cite{anthropic2025claudecode}      & \iconCmd+\iconDir+\iconNet  & \iconConfig+mode    & /tool+res      & \iconApp\,+\,\iconOS         & \iconPartial & \iconExpandOnly \\
    \rowcolor{rowSystem} \codex~\cite{openai2025codex}                    & \iconEnv+\iconNet           & mode sel.           & global         & \iconOS                      & \iconClosed  & \iconImmutable \\
    \rowcolor{rowSystem} \manus                                          & \iconEnv                    & auto VM             & /session       & \iconVM                      & \iconClosed  & \iconImmutable \\
    \rowcolor{rowSystem} OpenClaw~\cite{shan2026don}                & \iconTool+\iconNet          & JSON \iconConfig    & /tool+agent    & \iconFramework\,+\,\iconVM   & \iconTiered  & \iconFlexible \\
    \rowcolor{rowSystem} \cursorsys~\cite{cursor2025security}             & \iconCmd+\iconFile+\iconMCP & \iconConfig+\iconGUI & /tool          & \iconApp\,+\,\iconOS         & \iconPartial & \iconFlexible \\
    \rowcolor{rowSystem} Windsurf                                        & \iconDir+\iconNet           & \iconConfig+\iconAuth & /resource    & \iconApp\,+\,\iconCloud      & \iconPartial & \iconFlexible \\
    \rowcolor{rowSystem} \devin~\cite{devin2025cognition}                 & \iconEnv+\iconRepo          & \iconGUI            & /repo          & \iconVM                      & \iconClosed  & \iconExpandOnly \\
    \rowcolor{rowSystem} GitHub Copilot                                  & \iconCmd+\iconRepo+\iconNet & \iconConfig+mode    & /repo          & \iconOS                      & \iconPartial & \iconExpandOnly \\
    \rowcolor{rowSystem} Gemini CLI~\cite{google2025geminicli}            & \iconCmd+\iconNet           & TOML+mode           & /tool          & \iconOS                      & \iconClosed  & \iconExpandOnly \\
    \rowcolor{rowSystem} Copilot CLI~\cite{github2025copilotcli}          & \iconDir+\iconTool          & CLI flags           & /tool+res      & \iconApp                     & \iconClosed  & \iconExpandOnly \\
    \rowcolor{rowSystem} LangChain~\cite{langchain2025}                   & \iconTool                   & \iconCode           & /tool          & \iconFramework               & \iconOpen    & \iconFlexible \\
    \rowcolor{rowSystem} CrewAI~\cite{crewai2025}                         & \iconTool+deleg.            & \iconCode           & /tool (task)   & \iconFramework               & \iconOpen    & \iconFlexible \\
    \rowcolor{rowSystem} Bedrock~\cite{aws2025bedrock}                    & \iconAPI+\iconData          & console+IAM         & /resource      & \iconCloud                   & \iconTiered  & \iconFlexible \\
    \rowcolor{rowSystem} Google ADK~\cite{google2025adk}                  & \iconCode+\iconAPI          & \iconAuth+\iconEnv  & /OAuth         & \iconVM\,+\,\iconCloud       & \iconClosed  & \iconImmutable \\
    \rowcolor{rowSystem} MS Agent FW~\cite{microsoft2025agentfw}          & \iconAPI+\iconData          & \iconProxy          & /tool (role)   & \iconFramework               & \iconClosed  & \iconFlexible \\
    \rowcolor{rowSystem} \aider~\cite{aider2025}                          & ---                         & ---                 & ---            & \iconNone                    & \iconOpen    & --- \\
    \rowcolor{rowSystem} ChatGPT                                         & \iconAPI                    & GPT Builder         & /action        & \iconApp\,+\,\iconCloud      & \iconPartial & \iconImmutable \\
    \midrule
    \rowcolor{rowPlugin} mcp-guardian~\cite{mcpguardian2025}              & \iconMCP                    & policy \iconConfig  & /call          & \iconFramework               & \iconConfigurable & \iconFlexible \\
    \rowcolor{rowPlugin} Intercept~\cite{intercept2025}                   & \iconMCP                    & YAML                & /param         & \iconFramework               & \iconClosed  & \iconFlexible \\
    \rowcolor{rowPlugin} assay                                           & \iconMCP                    & policy              & /tool          & \iconFramework               & \iconClosed  & \iconFlexible \\
    \rowcolor{rowPlugin} GUARDRAIL                                       & \iconMCP+\iconAuth          & \iconConfig         & /tool          & \iconFramework               & \iconClosed  & \iconFlexible \\
    \rowcolor{rowPlugin} mcp-security-hub                                & \iconTool bundle            & install sel.        & /tool          & \iconApp                     & \iconOpen    & \iconFlexible \\
    \rowcolor{rowPlugin} cursor-security-rules                           & \iconCode-gen               & \texttt{.cursorrules} & global       & \iconApp                     & \iconOpen    & \iconFlexible \\
    \bottomrule
  \end{tabular}

  \vspace{4pt}
  \raggedright\small
  \textbf{Legend ---}
  Enforc.: \iconFramework\,framework, \iconOS\,OS, \iconVM\,VM, \iconCloud\,cloud, \iconApp\,application, \iconNone\,none.\quad
  Default: \iconOpen\,open, \iconClosed\,closed, \iconPartial\,partial, \iconTiered\,tiered, \iconConfigurable\,configurable.\quad
  Mutab.: \iconImmutable\,immutable, \iconExpandOnly\,expand-only, \iconFlexible\,flexible.\quad
  Entities: \iconCmd\,command, \iconTool\,tool, \iconMCP\,MCP, \iconAPI\,API, \iconFile\,file, \iconDir\,directory, \iconNet\,network, \iconEnv\,exec.\ env, \iconRepo\,repo, \iconCode\,code, \iconConfig\,config, \iconGUI\,GUI, \iconAuth\,OAuth, \iconData\,data, \iconProxy\,proxy.
\end{table*}

\noindent\textbf{Boundary object.}
Six classes of object are bounded across our corpus. Tools and commands are
bounded by \claudecode{}'s blocklist, OpenClaw's per-agent allow/deny lists,
and Copilot CLI's \texttt{--allow-tool}/\texttt{--deny-tool} flags. Files
and directories are bounded by Copilot CLI's directory trust model, Windsurf's
\texttt{.codeiumignore}, \cursorsys{} dotfile protection, and \claudecode{}'s
sandbox directory restrictions. Networks are bounded by \codex{}'s
network-off default, OpenClaw's Docker network control, and \claudecode{}
sandbox per-host allowlists. Execution environments are bounded by
\manus{}'s isolated cloud VM, \devin{}'s sandboxed shell-editor-browser
environment, and Google ADK's gVisor runtime~\cite{google2025adk}. Code
repositories or cloud resources are bounded by \devin{}'s repository access,
GitHub Copilot's repo scope, and Bedrock guardrails' IAM resource scope.
APIs and credentials are bounded by Google ADK OAuth scopes,
Bedrock's read-only database recommendations, and \devin{}'s secrets
manager.

\noindent\textbf{Boundary definition.}
How the boundary is articulated. CLI flags are the most lightweight definition
(Copilot CLI \texttt{--allow-tool}/\texttt{--deny-tool}, \codex{}'s sandbox-mode
flag). Configuration files are the modal production form (OpenClaw JSON,
Gemini CLI TOML, Windsurf \texttt{.codeiumignore}, \claudecode{}
\texttt{settings.json}). GUI or admin consoles are used in enterprise
contexts (Bedrock AWS console, \cursorsys{} admin settings, \devin{}
repository selector). Mode selection from a discrete set of safety profiles
(\codex{} \texttt{workspace-write} versus \texttt{danger-full-access},
\claudecode{} sandbox on/off) is the simplest user-facing form. Natural
language definitions in \texttt{CLAUDE.md} or \cursorsys{} rules files are
cheap to write but fall back to LLM self-compliance for enforcement.
OAuth or IAM definitions (Google ADK OAuth scopes, Bedrock IAM condition keys,
Windsurf SSO/SCIM) delegate the boundary to an existing identity layer.

\noindent\textbf{Granularity.}
The granularity of the boundary itself. Global mode, where a single switch
governs all behavior (\codex{} sandbox modes, \claudecode{} sandbox on/off,
Copilot CLI YOLO mode), is the coarsest. Per-resource granularity configures
each directory, repo, or network host separately (Copilot CLI per-directory
trust, \devin{} per-repo access, \claudecode{} per-host network allowlist).
Per-tool granularity gives each tool its own allow or deny entry
(\claudecode{} blocklist, OpenClaw allow/deny, Copilot CLI per-tool flags).
Per-parameter granularity, the finest in our corpus, distinguishes parameter
values within a single tool call (Gemini CLI's TOML engine, Intercept's YAML
rules).

\noindent\textbf{Enforcement level.}
The layer of the stack that mediates the boundary. Application-level
enforcement (\claudecode{} blocklist, \cursorsys{} dotfile protection,
ChatGPT GPT configuration) has the lowest bypass difficulty: an application
bug or successful prompt injection can defeat the check. Framework or
middleware enforcement (LangChain scoped tool access, CrewAI task-tool
scoping, OpenClaw three-layer permission gates, plus all four MCP-proxy
plugins) sits one rung above. OS-level enforcement (\claudecode{} macOS
Seatbelt and Linux Bubblewrap, \codex{} OS-native sandbox, \cursorsys{}
OS-level agent sandboxing, Gemini CLI Seatbelt or container) requires an
OS-level vulnerability to bypass. VM or container isolation (\manus{} cloud
VM, OpenClaw Docker, Google ADK gVisor, \devin{} sandbox) requires a
container escape. Cloud-platform enforcement (Bedrock IAM with guardrail
identifiers, Google ADK Vertex AI Agent Engine, Windsurf self-hosted) sits
at the highest bypass-difficulty point.

\noindent\textbf{Default stance.}
What the system does before any user configuration. Default-open, where
all actions are allowed and the user must lock down (\aider{}'s no-sandbox
default~\cite{aider2025}, Copilot CLI YOLO mode, \cursorsys{} Auto-Run), is
frictionless but is the source of every public incident in our corpus,
including Replit Agent's July 2025 production database
deletion~\cite{replit2025agent}. Default-partial (\claudecode{}'s default
ask mode, \cursorsys{}'s default mix of automatic file edits with prompted
shell commands) gates dangerous actions but allows common ones. Default-closed,
where all actions are denied until configured (\codex{}'s default
\texttt{workspace-write}, Gemini CLI's default restricted, Copilot CLI's
per-directory trust), is the safest default but with the highest configuration
burden. Tiered default (OpenClaw defaults the agent layer to deny while
leaving sandbox-internal tools allowed; Bedrock requires explicit guardrail
association on top of inherited IAM defaults) splits the default across
layers.

\noindent\textbf{Mutability.}
How the boundary changes during the session. Immutable boundaries fixed at
startup (\manus{} cloud VMs, \codex{} sandbox modes once chosen, Docker
container network isolation) are the strongest property: scope cannot drift.
Expand-only mutability, where the runtime allows widening but not narrowing,
is the dominant production sub-type (\claudecode{}'s ``always allow''
history, Gemini CLI's permission expansion dialog, Copilot CLI's accumulated
session-wide permissions, \devin{}'s escalating capability grants); we name
this pattern \emph{scope creep} elsewhere in the paper. Flexible bidirectional
mutability (\cursorsys{} admin policy, OpenClaw configuration changes,
\claudecode{} settings edits) supports narrowing in principle, although in
practice users rarely shrink boundaries that have already been expanded.

\noindent\textbf{Cross-cutting observations.}
C4 is the category where academia and industry are closest in spirit but
most different in implementation level. Academic proposals formalize
isolation at the framework or design layer (IsolateGPT,
MiniScope~\cite{zhu2025miniscope},
FaultTolerantSandbox~\cite{yan2025fault},
AgentBound), while production systems operationalize the same idea at every
available layer of the stack: application-level lists, OS-level sandboxes,
VM-level isolation, and cloud-platform IAM. The diversity of enforcement
layers is itself a finding: a single production system typically stacks two
or three layers (\claudecode{} stacks application blocklist with OS sandbox;
OpenClaw stacks framework with Docker; Google ADK stacks gVisor with cloud
IAM). A second pattern is the asymmetry between configuration and
reconfiguration: default-closed systems pay the cost up front but are also
where users most often relax the configuration in ways they later forget,
while default-partial and default-open systems rely entirely on the
runtime-approval system to backstop the omission. The expand-only mutability
that dominates production then converts every backstop click into a
permanent surface widening, and few production systems we surveyed offer any
visualization of how the trusted surface has evolved during a session.

\takeaways{\label{tk:app-c4-stack}%
Production C4 instantiations stack two or three enforcement layers per
system, but the layering does not address mutability: the dominant runtime
behavior is expand-only, with no production system offering a session-level
view of accumulated scope expansions. The default-stance choice (open,
partial, closed) is the only knob that meaningfully shifts the up-front
configuration burden, and even default-closed systems drift toward open as
the session accrues approvals.
}

\subsection{C5: Intent Anchoring}
\label{app:c5}

Intent anchoring is the only category where the user takes no explicit
security action: the system treats the user's normal task description, query
history, or staging-phase behavior as a security ground truth and audits
agent execution against it. Across our corpus, 5 academic papers ship intent
anchoring designs but \emph{zero} production systems and \emph{zero}
plugins do. We organize the design space along four dimensions: the source
of the anchor, the deviation-detection method, the timing of detection, and
the user's awareness of the mechanism. Table~\ref{tab:c5-dims} reports the
full coding.

\begin{table*}[t]
  \centering
  \small
  \caption{Intent Anchoring: every paper in our corpus that ships an intent-anchoring mechanism, characterized along four dimensions. No production system or plugin currently implements intent anchoring, so the table contains academic sources only (rows in \colorbox{rowPaper}{\strut blue}). The \emph{User awareness} column highlights that the user never performs an explicit security action---this is the defining property of implicit \ahi{}.}
  \label{tab:c5-dims}
  \setlength{\tabcolsep}{4pt}
  \begin{tabular}{@{}l c c c c@{}}
    \toprule
    \textbf{Source} & \textbf{Anchor source} & \textbf{Detection} & \textbf{Timing} & \textbf{User aware.} \\
    \midrule
    \rowcolor{rowPaper} TaskShield~\cite{jia2025task}         & \iconSinglePrompt  & \iconSemAlign   & \iconPerStep     & \iconInformed \\
    \rowcolor{rowPaper} DRIFT~\cite{li2025drift}                    & \iconDynRules      & \iconRuleMatch  & \iconPerStepCont & \iconInformed \\
    \rowcolor{rowPaper} MELON~\cite{zhu2025melon}                   & \iconSinglePrompt  & \iconReExec     & \iconBatch       & \iconNoAware \\
    \rowcolor{rowPaper} Agent-Sentry~\cite{sequeira2026agent}         & \iconSinglePrompt  & \iconSemAlign   & \iconPerStep     & \iconIndirect \\
    \rowcolor{rowPaper} AgentGuardian~\cite{abaev2026agentguardian} & \iconBehavHist     & \iconBehavAnom  & \iconContinuous  & \iconNoAware \\
    \bottomrule
  \end{tabular}

  \vspace{4pt}
  \raggedright\small
  \textbf{Legend ---}
  Anchor: \iconSinglePrompt\,user prompt, \iconDynRules\,query+rules, \iconBehavHist\,behav.\ history.\quad
  Detection: \iconSemAlign\,semantic align., \iconRuleMatch\,rule match, \iconReExec\,re-exec.\ compare, \iconBehavAnom\,behav.\ anomaly.\quad
  Timing: \iconPerStep\,per-step, \iconPerStepCont\,per-step+cont., \iconBatch\,batch, \iconContinuous\,continuous.\quad
  Awareness: \iconInformed\,informed (no action), \iconNoAware\,none, \iconIndirect\,indirect.
\end{table*}

\noindent\textbf{Anchor source.}
Where the trusted reference comes from. The single user prompt is the
modal anchor, used by 3 of 5 papers:
TaskShield~\cite{jia2025task} treats the user's task specification as
ground truth, MELON~\cite{zhu2025melon} uses the prompt as the masking
reference, and Agent-Sentry~\cite{sequeira2026agent} anchors provenance audits
on the original request. Query plus dynamically derived rules, where security
rules are compiled from the query at task time, appears in
DRIFT~\cite{li2025drift}. Behavioral history, where normal behavior is
learned from a staging phase or rolling history, appears in
AgentGuardian~\cite{abaev2026agentguardian}.

\noindent\textbf{Deviation detection method.}
How the system decides whether an action is consistent with the anchor.
Semantic alignment checks, where an LLM judges whether a candidate action
serves the declared task, appear in TaskShield (2.07\% attack success rate)
and Agent-Sentry (90\%+ detection). Re-execution comparison
(MELON's masked re-execution) provides a provable defense, since divergence
between masked and unmasked runs is direct evidence of injection. Rule
matching with dynamic update, where synthesized rules are matched
deterministically and updated as context shifts, is DRIFT's approach,
combined with memory-stream isolation. Behavioral anomaly detection, where
runtime behavior is compared against a learned baseline, is AgentGuardian's
approach.

\noindent\textbf{Detection timing.}
When the check runs relative to the action. Per-step timing, where each
action is validated before or after execution, is the modal choice
(TaskShield, DRIFT, Agent-Sentry). Batch timing, where detection runs over
a sequence of actions rather than at each step, applies to MELON's masked
re-execution which requires enough trajectory to compare. Continuous
learning, where the detection baseline is updated as the session proceeds,
applies to AgentGuardian and to DRIFT (which also updates its rules during
execution).

\noindent\textbf{User awareness.}
How visible the security mechanism is to the user. Fully invisible
mechanisms, where detection runs in the background and the user has no idea
their prompt is being used as a security anchor, are the modal choice
(MELON, AgentGuardian). Indirect awareness, where the user knows there is a
security mechanism but not that their prompt is the anchor, applies to
Agent-Sentry: its user-facing surface mentions security checks without
exposing the provenance audit. Informed-but-no-action mechanisms tell the
user that the prompt anchors the security check while requiring nothing
extra (TaskShield notes that vague prompts reduce defensive efficacy;
DRIFT surfaces the synthesized rules).

\noindent\textbf{Cross-cutting observations.}
The most striking pattern in C5 is that user-burden cost cannot explain the
zero industry adoption. Unlike C2, where the burden is the canonical
hypothesis, C5's user burden is exactly zero by construction: the user only
writes the prompt they would have written anyway. Two alternative hypotheses
are consistent with what we observe. First, invisibility makes intent
anchoring difficult to market: a user who never sees the safeguard does not
perceive new value. Second, the false-positive mode is unusually hostile: an
agent that refuses to do something the user just asked it to do, with no
obvious trigger, is harder to recover from than an approval prompt. Neither
hypothesis has been empirically tested in our corpus. A second observation
is that C5 overlaps non-trivially with C1 and C3: DRIFT compiles security
rules from the user query, an instance of LLM-generated dynamic policy that
is also classifiable as C1; TaskShield's per-step semantic check resembles
a runtime approval where the approver is the LLM judge rather than the
human, blurring the C5/C3 boundary. The category remains distinct because
the user takes no security action, but the implementation overlaps with
mechanisms in better-adopted categories, which makes the persistent
zero-adoption result more puzzling, not less.

\takeaways{\label{tk:app-c5-invisibility}%
Intent anchoring has zero user-facing burden by construction, so the
cost-based hypothesis used to explain C2's gap does not apply. The
remaining hypotheses, namely invisibility (no perceived value) and
false-positive attribution (hard to explain a refusal that has no
prompt), are consistent with the observed data but untested. The fact
that C5 mechanisms overlap with C1 and C3 patterns that \emph{are} adopted
makes the gap more striking, not less.
}

\subsection{Cross-Category Observations}
\label{app:cross}

Having walked through the five categories individually, we now compare
them on adoption, composition, and the security-usability frontier.
Table~\ref{tab:cross-cat} reports the cross-category counts.

\begin{table*}[t]
  \centering
  \small
  \caption{Cross-category coverage across \npapers academic papers,
  \nsystems production agent systems, and \nplugins security plugins. For
  each category we report the count of sources providing a mechanism in
  that category and list the predominant enforcement layer(s) observed.
  Subsections are presented in adoption-descending order (runtime approval
  first).}
  \label{tab:cross-cat}
  \setlength{\tabcolsep}{3pt}
  \begin{tabular}{@{}l cccc >{\raggedright\arraybackslash}p{5.5cm}@{}}
    \toprule
    \textbf{Category} &
    \textbf{Papers} &
    \textbf{Systems} &
    \textbf{Plugins} &
    \textbf{Total} &
    \textbf{Predominant enforcement} \\
    \midrule
    Runtime Approval
      & 5 & \textbf{15/\nsystems} & 4 & 24
      & Application / framework proxy \\
    Scope \& Boundary Config.
      & 4 & \textbf{16/\nsystems} & 6 & 26
      & Application $\to$ OS $\to$ VM $\to$ cloud \\
    Policy Specification
      & 13 & 14/\nsystems & 8 & 35
      & NL self-compl.\ (industry); det.\ DSL (academia) \\
    Trust \& Data Labeling
      & 8 & \textbf{0/\nsystems} & 0 & 8
      & IFC / capability (academia only) \\
    Intent Anchoring
      & 5 & \textbf{0/\nsystems} & 0 & 5
      & Semantic align.\ / program analysis (academia only) \\
    \bottomrule
  \end{tabular}
\end{table*}

\noindent\textbf{Coverage by category.}
The five categories span a wide adoption range. C3 and C4 are near-universal
in production (15 and 16 of 21 systems respectively) and account for the bulk
of plugin offerings. C1 is heavily represented in academia (13 papers) and
widely adopted in production (14 systems), but under different
representations: academia favors deterministic DSLs and permission models,
while production overwhelmingly favors natural-language and declarative-config
policies. C2 and C5 together account for 13 academic papers and zero
industry adoption, the two largest gaps in our corpus.

\noindent\textbf{Academia-industry gap.}
The gap is not uniform. In C2, the gap is total: every academic mechanism
(full IFC, capability labels, default-deny capability checks, selective
propagation, formal verification) lacks a production analogue. In C5, the
gap is similarly total, but the underlying obstacle differs: C2's burden
hypothesis is contradicted by the failure of low-burden variants to gain
adoption, while C5's burden is structurally zero. In C1, the gap is
qualitative: academia and industry both ship policy specification mechanisms
but disagree on representation and enforcement. In C3 and C4, the gap is
smallest: academic proposals refine the same mechanisms industry already
deploys, with the notable exception of risk-adaptive approval (RTBAS), which
has zero industry adoption despite clear utility-security advantages.

\noindent\textbf{Security-usability tradeoffs.}
Projecting the 21 production systems onto a (security, usability) plane
reveals three clusters. High-security, high-burden systems default closed
and ask frequently: \codex{}, Gemini CLI, and Bedrock fall here.
Low-security, low-burden systems default open: \aider{} and the early Replit
Agent fall here, and both have associated incidents. Middle-band systems
combine runtime approval with always-allow memory: \claudecode{},
\cursorsys{}, Copilot CLI, \manus{}, and most other mainstream agents
cluster in this band. The high-security low-burden quadrant is occupied
only by academic proposals: RTBAS (C3 risk-adaptive), TaskShield (C5 intent
anchoring), CaMeL (C2 capability), and Conseca (C1 dynamic policy). The same
mechanisms that could in principle dominate the desirable quadrant are the
ones with zero production adoption.

\noindent\textbf{Multi-category composition.}
Mature production systems do not pick a single category; they stack three or
four. \claudecode{} combines C1 (\texttt{CLAUDE.md} plus Hooks plus blocklist),
C3 (per-command approval), and C4 (sandbox plus blocklist plus network
allowlist). \codex{} combines C1 (rules), C3 (layered approval), and C4
(sandbox modes). Gemini CLI combines C1 (TOML policy engine), C3 (default
restricted), and C4 (OS sandbox). The C3 plus C4 combination forms a
complementary pair: fatigue is mitigated by widening the sandbox (so
approvals fire less often), while sandbox misconfiguration is backstopped by
approval prompts. The pair inherits the weaknesses of both: silent scope
creep from C3 plus expand-only mutability from C4. We did not find a single
production system that combines C2 or C5 with the dominant C1+C3+C4 stack,
despite the natural complementarity.

\takeaways{\label{tk:app-cross-frontier}%
The (security, usability) frontier is occupied by academic proposals in
the desirable quadrant and by production systems in the middle band. The
production C1+C3+C4 stack inherits the weaknesses of both C3 (silent
scope creep) and C4 (expand-only mutability). The two categories that
could in principle close the gap (C2 and C5) are absent from every
production stack we surveyed.
}

\section{Research Agenda: Extended Sub-Directions}
\label{app:agenda-extended}

The main text presents two key sub-directions per research direction. Here we
describe additional promising avenues.

\subsection{Direction 1 (Extended): Cognitive Load-Aware Design}

Beyond risk-adaptive approval and bidirectional scope adjustment, a third
sub-direction draws on the usable-security literature: \emph{cognitive
load-aware interaction design}. This approach measures and models the user's
real-time cognitive state (through interaction patterns, response times, and
task complexity) to dynamically adjust the frequency, detail level, and
urgency framing of security prompts. The goal is to ensure that security
decisions are presented when the user has the cognitive bandwidth to make
them thoughtfully, rather than at arbitrary runtime moments dictated by the
agent's execution sequence.

\subsection{Direction 2 (Extended): Template-Based Policy Authoring}

Natural-language policies are easy to write but unenforceable; formal DSLs are
enforceable but inaccessible. \emph{Template-based policy authoring} offers a
middle path: the system provides a library of pre-built policy templates
(e.g., ``read-only file access,'' ``network requests to approved domains only'')
that users select and parameterize. Templates are backed by formal semantics,
so enforcement is deterministic, but the user experience is closer to
configuring preferences than writing code. Template libraries can be extended
by security experts and shared across organizations.

\subsection{Direction 3 (Extended): Preference Learning and Structured Elicitation}

Two additional sub-directions complement LLM-generated policy drafts and
proactive ambiguity detection.
\emph{Preference learning from approval histories} trains lightweight models
on a user's past approve/deny decisions to predict future preferences,
progressively reducing interruption frequency as the system learns which
action classes the user consistently approves. Unlike ``always allow'' (which
is action-specific and permanent), preference learning generalizes across
similar actions and can decay over time.
\emph{Structured intent elicitation} uses LLM-guided dialogues to help users
articulate their intent in a structured format, analogous to IDE
auto-complete for security preferences. Rather than extracting intent from
a freeform prompt after the fact, the system prompts the user with targeted
questions (``Should the agent be allowed to access files outside the project
directory?'') at task specification time, building a structured intent
profile that downstream security mechanisms can enforce deterministically.


\end{document}